\documentclass[aps,prl,twocolumn,superscriptaddress,nofootinbib]{revtex4-2}
\usepackage{graphicx,amssymb,amsmath,amsfonts,empheq,latexsym,braket,bm,bbm,dsfont,upgreek}
\usepackage[normalem]{ulem}
\usepackage[dvipsnames]{xcolor}
\usepackage{comment}
\usepackage{tabularx}
\newcolumntype{C}{>{\centering\arraybackslash}X}
\usepackage{color}
\usepackage[dvipsnames]{xcolor}
\usepackage{makecell}
\usepackage{multirow}
\usepackage{url}
\usepackage{soul}
\usepackage{algorithm}
\usepackage[percent]{overpic} 
\usepackage{algpseudocode}
\usepackage{hyperref}
\usepackage{svg}
\hypersetup{
colorlinks = true,
linkcolor = [rgb]{0,0,0}, 
citecolor = [rgb]{0.13,0.55,0.13},
urlcolor = [rgb]{0.25, 0.41, 0.88}}

\usepackage{tikz}
\usepackage{tikz-cd}
\usetikzlibrary{arrows}
\usetikzlibrary{intersections}
\usetikzlibrary{shapes.geometric}
\usetikzlibrary{decorations.pathmorphing, patterns,shapes}
\usetikzlibrary{decorations.markings}

\tikzset{
	mid arrow/.style={postaction={decorate,decoration={
				markings,
				mark=at position .575 with {\arrow[#1]{stealth}}
	}}},
	near arrow/.style={postaction={decorate,decoration={
				markings,
				mark=at position .275 with {\arrow[#1]{stealth}}
	}}},
	far arrow/.style={postaction={decorate,decoration={
				markings,
				mark=at position .800 with {\arrow[#1]{stealth}}
	}}},
}

\renewcommand{\leq}{\leqslant}

\newcommand{\const}{\operatorname{const}}

\newcommand{\bbR}{\mathbb{R}}

\newcommand{\bbZ}{\mathbb{Z}}

\renewcommand{\r}{\boldsymbol{r}}

\newcommand{\eqnref}[1]{Eq.~\eqref{#1}}
\newcommand{\figref}[1]{Fig.~\ref{#1}}

\begin{document}

\title{Conformal scalar field theory from Ising tricriticality on the fuzzy sphere}

\author{Joseph Taylor}
\affiliation{School of Physics and Astronomy, University of Leeds, Leeds LS2 9JT, United Kingdom}
\author{Cristian Voinea}
\affiliation{School of Physics and Astronomy, University of Leeds, Leeds LS2 9JT, United Kingdom}
\author{Zlatko Papi\'c}
\affiliation{School of Physics and Astronomy, University of Leeds, Leeds LS2 9JT, United Kingdom}
\author{Ruihua Fan}
\affiliation{Department of Physics, University of California, Berkeley, CA 94720, USA}

\begin{abstract}

Free theories are landmarks in the landscape of quantum field theories: their exact solvability serves as a pillar for perturbative constructions of interacting theories. 
Fuzzy sphere regularization, which combines quantum Hall physics with state-operator correspondence, has recently been proposed as a promising framework for simulating three-dimensional conformal field theories (CFTs), but so far it has not provided access to free theories. We overcome this limitation by designing a bilayer quantum Hall system  that hosts an Ising tricritical point—a nontrivial fixed point where first-order and second-order transitions meet—which flows to the conformally coupled scalar theory in the infrared. The critical energy spectrum and operator structure match those at the Gaussian fixed point, providing nonperturbative evidence for the emergence of a free scalar CFT. 
Our results expand the landscape of CFTs realizable on the fuzzy sphere and demonstrate that even free bosonic theories – previously inaccessible – can emerge from interacting electrons in this framework.

\end{abstract}

\maketitle

{\bf \em Introduction.---}Understanding quantum field theories (QFTs) in dimensions higher than two remains a central challenge in theoretical physics, especially in the conformally-invariant regime where perturbative methods often fail~\cite{Rychkov:2016iqz,Simmons-Duffin:2016gjk,Poland19}. The fuzzy sphere regularization, which combines state-operator correspondence~\cite{Cardy84} with insights from quantum Hall physics~\cite{Prange87}, has recently emerged as a powerful framework for simulating three-dimensional (3D) conformal field theories (CFTs)~\cite{Zhu23}. 
This approach enables the computation of operator-product expansion (OPE) coefficients and correlators~\cite{HuOPE2023,Han:2023yyb}, conformal generators~\cite{fardelli2024constructinginfraredconformalgenerators,fan2024noteexplicitconstructionconformal}, entropic $F$-functions~\cite{hu2025entropic}, and study of defects and surface CFTs~\cite{Hu2024,Zhou2024,Zhou2025,dedushenko2024isingbcftfuzzyhemisphere}.
It has successfully captured a range of interacting theories, including the 3D Ising and $O(N)$ Wilson-Fisher models~\cite{Zhu23,Han24,voinea2024regularizing, Lauchli25}, deconfined quantum critical points~\cite{Ippoliti:2018prb,Zhou:2024qfi,chen24a,chen24b}, the pseudocritical Potts model~\cite{yang2025microscopic}, non-unitary Yang-Lee models~\cite{fan2025simulatingnonunitaryyangleeconformal,cruz2025yangleequantumcriticalityvarious,miro2025flowingisingmodelfuzzy} and other new classes of CFTs~\cite{Zhou24b}.

One glaring exception in the set of theories realized on the fuzzy sphere are \emph{free} CFTs. As exactly solvable models, free CFTs are reference points for elucidating the fundamental structural features of QFTs and analyzing strongly interacting systems.
Therefore, they play a foundational role in QFT but many-body physics more broadly.
The realization of free theories on the fuzzy sphere would not only be a step towards establishing the completeness of the fuzzy sphere regularization but it would also offer conceptual insights into its mechanism.

In this work, we show how to realize the simplest free CFT -- the conformally coupled scalar, which we refer to as the free scalar throughout this paper. At first glance, it may seem unlikely that a free bosonic theory could emerge from a system of strongly-interacting electrons in the lowest Landau level (LLL), which the fuzzy sphere method utilizes.
The key insight comes from the nature of Ising tricriticality, described by the Euclidean Lagrangian, $\mathcal{L}_E = (\partial \phi)^2 + g \phi^6$, in flat space.
In 3D, the $\phi^6$ interaction is marginally irrelevant at the Gaussian fixed point, so a system at tricriticality will flow toward the free scalar CFT in the deep infrared (IR)~\cite{cardy1996scaling,henriksson2025tricriticalisingcftconformal}.
This perspective provides a natural route to the free scalar: tune a quantum Hall Ising ferromagnetic transition to its tricritical point.

Specifically, we design a bilayer quantum Hall system that realizes both an Ising transition line and a first-order transition line, and most importantly, the Ising tricriticality at their intersection point. Our two-component model is conceptually different from the standard one of tricriticality based on spin-1 degrees of freedom~\cite{Blume1966,Capel1966}. With appropriate fine-tuning, our model at the tricritical point is characterized by a free scalar CFT up to small residual interactions, which we demonstrate by the matching of the low-energy spectra and the emergence of the characteristic free-boson algebra.

{\bf \em The model.---}  
Consider a quantum Hall bilayer on a sphere~\cite{Haldane:1983xm}. 
Setting the filling factor and magnetic length to unity, the number of flux quanta $N_\phi$, the number of electrons $N$ and the radius $R$ are related by $N_\phi = N {-} 1 = 2R^2$.
Let $c_{a = \uparrow,\downarrow}(\r)$ denote the LLL-projected electron annihilation operators for the two layers in real space. 
The Hamiltonian reads
\begin{eqnarray}\label{eq:TCI fuzzy}
	\nonumber H &=&  H_{\text{inter}} + \lambda H_{\text{intra}} - h\int  \left( c_\uparrow^\dagger(\r)c_\downarrow(\r) + \mathrm{h.c.} \right)\, d^2 \r\,, \\
    H_{\text{inter}} &=& 2\int V(\r_1{-}\r_2) \, n_\uparrow (\r_1) n_\downarrow(\r_2) \, d^2 \r_1 d^2 \r_2  \,, \\
	\nonumber H_{\text{intra}} &=& \sum_{a = \uparrow,\downarrow} \int U(\r_1{-}\r_2) \, :n_a (\r_1) n_a (\r_2): d^2 \r_1 d^2 \r_2 \,,   
\end{eqnarray}
where $n_a(\r) = c_a^\dag(\r) c_a(\r)$ are the layer-pseudospin operators and $:\;:$ denotes normal ordering. The intra- and inter-layer interactions, $U$ and $V$, to be specified below, are assumed to be predominantly repulsive and we parametrize them by the Haldane pseudopotentials, $U_J$ and $V_J$~\cite{Haldane:1983xm}. We keep charge degrees of freedom gapped and let the pseudospins undergo a transition.

Compared to the 3D Ising fuzzy sphere model~\cite{Zhu23}, our Hamiltonian \eqref{eq:TCI fuzzy} contains an extra intralayer interaction term, which plays a crucial role in realizing the first-order transition. This can be understood at the mean-field level via a Hartree–Fock approximation.
As shown in the Supplementary Material (SM)~\cite{SM}, the phase boundary is given by $h = 2 (1 - \lambda) \sum_J (V_{2J+1} - U_{2J+1})$. 
Fixing $V$ and $U$, the system is ferromagnetic at small values of $\lambda$ and $h$, and paramagnetic otherwise. 
The transition is second-order for all $h > 0$ and becomes first-order only at $h = 0$.
Although the mean-field analysis does not capture the tricritical point on the phase boundary, it highlights the necessity of including intralayer interactions to realize first-order behavior.
Below we will see that fluctuations beyond mean field qualitatively modify the critical behavior near $h = 0$, giving rise to richer phase structures that \emph{do} contain a tricritical point.

{\bf \em Phase diagram and tricritical point.---}
Unlike in two dimensions, where imposing the Kramers-Wannier duality reduces the number of $\bbZ_2$-even relevant perturbations from two to one, no such counterpart exists in 3D~\cite{Grover:2013rc, Rahmani:2015, OBrien:2017wmx}. As a result, reaching the tricritical point requires careful tuning of parameters. Moreover, because the equivalence to the free scalar theory emerges only asymptotically in the IR, fine tuning is necessary to suppress finite-size effects, particularly those stemming from the $\phi^6$ interactions.

In our case, we first fix $\lambda = 1$ and use gradient descent in the residual parameter space to identify a point whose energy spectrum closely matches that of the free scalar.
We then perform two consistency checks: (1) vary $\lambda$ and $h$ to map out the full phase diagram and confirm that the identified point is at a tricritical juncture, and (2) explicitly demonstrate that the spectrum at this point is well-approximated by a free scalar CFT, even in system sizes beyond those for which the optimization was performed. We use gradient descent optimization over five parameters: two intralayer pseudopotentials $(U_3, U_5)$, two interlayer ones $(V_0, V_2)$, and the transverse field $h$, with fixed $U_1=V_1=1$ to avoid overfitting and accidental symmetries. Importantly, we also impose an additional constraint to avoid spurious local minima, where the spectrum seems conformal but the ground state wave function is actually gapped~\cite{SM}. Performing the described optimization for $N=10$, we arrive at $\{ U_1,U_3,U_5 \} \approx \{1, 0.07, -0.13 \}\,,\{ V_{0},V_{1},V_{2} \} \approx \{2.35, 1, 0.37 \}\,,  h \approx 0.23$.
With these parameters, the low-energy spectra for different system sizes match the free scalar theory, as we discuss in detail below, and the energy gap vanishes as $1/\sqrt{N}$, confirming the gapless nature of this point~\cite{SM}.

\begin{figure}[t]
\centering
\includegraphics[width=0.98\linewidth]{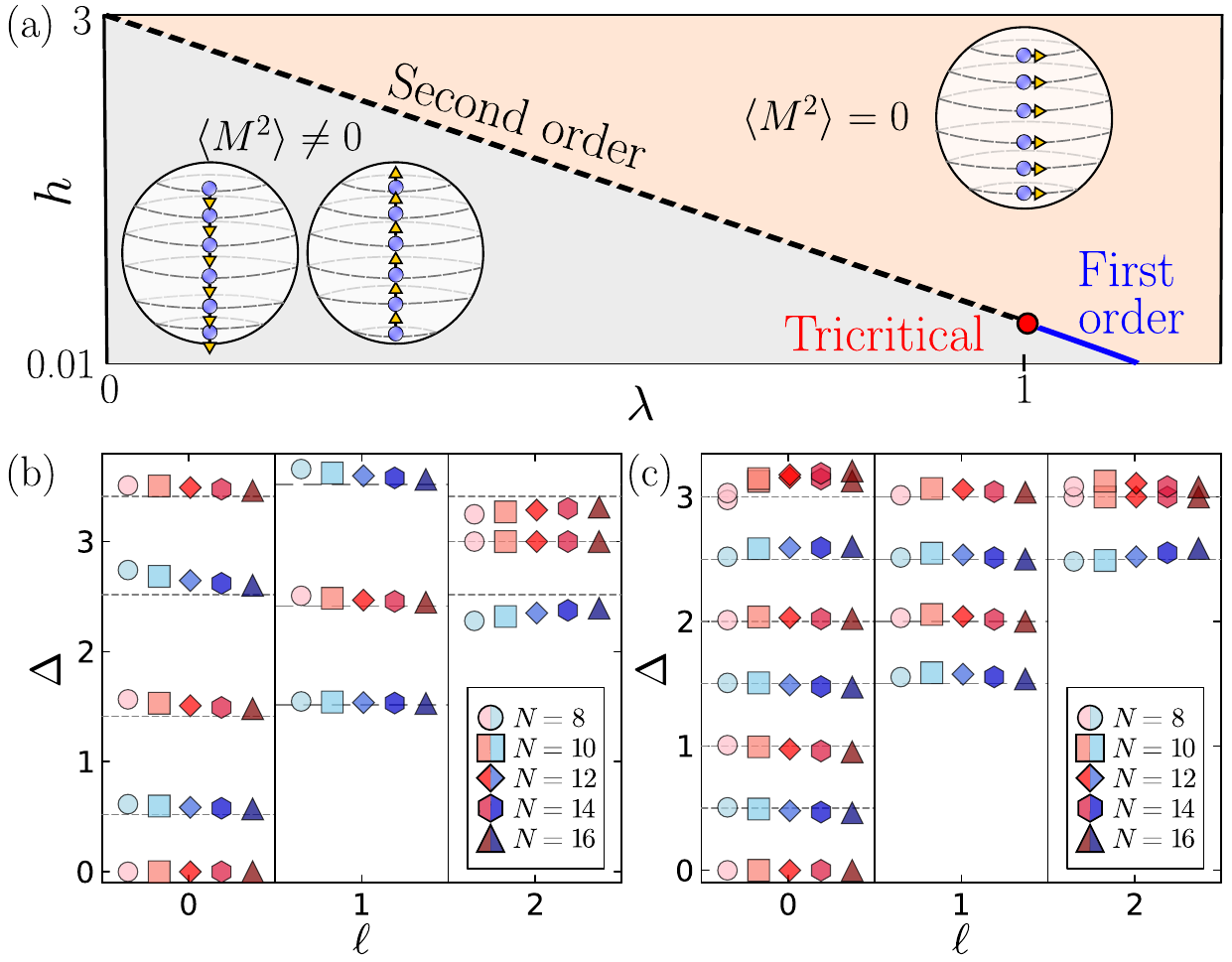}
\caption{Phase diagram and representative energy spectra. (a) The line of 3D Ising transitions (black dashed line) meets the line of first-order transitions (blue line) at a tricritical point  $(\lambda,h) \approx (1.0,0.23)$ (red dot). The shaded regions correspond to ordered and disordered phases according to the value of the magnetization, $\langle M^2\rangle$. 
(b) The energy spectrum as a function of angular momentum $\ell$  at a representative point $(\lambda, h) = (0.4, 2)$ along the 3D Ising line.  Data is for several system sizes shown in the legend, while red and blue symbols indicate $\mathbb{Z}_2$-even and odd states, respectively. We rescale the spectrum so that the first spin-$2$, $\mathbb{Z}_2$-even state lies at energy $3$. (c) Same as (b) but taken at the tricritical point.
While the spectrum in (b) shows good agreement with 3D Ising CFT~\cite{Zhu23}, the spectrum in (c) is consistent with free scalar theory, with (half-)integral energies alternating in $\mathbb{Z}_2$ parity. 
} 
\label{fig:phase diagram}
\end{figure}

Fixing the optimized parameters, we vary the overall intralayer interaction strength $\lambda$ and transverse field $h$ to map out the two-dimensional phase diagram shown in \figref{fig:phase diagram}(a). 
At large $h$, mean-field analysis predicts a second-order transition, which is expected to fall into the Ising universality class upon including fluctuations. Here, we estimate the location of the Ising critical line by leveraging state-operator correspondence. 
Specifically, we define a cost function that compares the low-energy spectrum at each $(\lambda, h)$ to that of the 3D Ising CFT.
At a fixed system size $N=12$, we find the cost function exhibits a minimum along a line in parameter space~\cite{SM}, which defines the black dashed line in \figref{fig:phase diagram}(a). Following this line towards smaller $h$, we find the ground state energy density as a function of $\lambda$ starts to develop a kink, suggesting that continuous phase transition gives way to a first-order line [blue line in \figref{fig:phase diagram}(a)], with the optimized parameter sitting in between as the putative tricritical point [red dot in \figref{fig:phase diagram}(a)].

Before presenting the supporting numerical analysis of the phase diagram, we point out one subtlety at $h=0$.
In this regime, the particle number in each layer is conserved and increasing $\lambda$ leads to a first-order transition at $\lambda\gtrsim 1$, from one of the layers being fully filled, $\nu_\uparrow{=}1$, $\nu_\downarrow{=}0$ (or vice versa) to $(\nu_\uparrow,\nu_\downarrow)=(1/2,1/2)$~\cite{SM}. 
While this occurs as a level crossing, consistent with our \figref{fig:phase diagram}(a), the state $(\nu_\uparrow,\nu_\downarrow)=(1/2,1/2)$ describes two charge-gapless composite fermion Fermi seas rather than a gapped paramagnet~\cite{Halperin93}. To avoid this, our phase diagram in \figref{fig:phase diagram}(a) starts from a small but non-zero value of $h=0.01$, which is sufficient to open up a gap near $\lambda=1$~\cite{SM}.

Along the dashed black line in \figref{fig:phase diagram}(a), the low-energy spectrum closely matches that of the 3D Ising CFT~\cite{Zhu23}. 
\figref{fig:phase diagram}(b) shows the spectrum at a representative point $(\lambda,h) = (0.40, 2.0)$, whose location is refined via finite-size scaling beyond the system size for the optimization. 

By contrast, the energy spectrum at the optimized parameter values 
closely matches that of the free scalar CFT, as shown in Fig.~\ref{fig:phase diagram}(c).  It is easiest to verify this matching via the state-operator correspondence. 
Recall the Euclidean Lagrangian of the free scalar $\mathcal{L}_E = (\partial \phi)^2/2$ in $\mathbb{R}^3$.
The field $\phi$ is a scalar primary with the scaling dimension $\Delta = 1/2$.
All other local operators are built from powers of $\phi$ and its derivatives, such as $\phi^n$, $\partial_\mu\phi^n$, etc., which thus have integral or half-integral scaling dimensions~\cite{SM}. This is indeed consistent with Fig.~\ref{fig:phase diagram}(c). A more detailed inspection reveals additional fingerprints of the free scalar theory. For example, since $\phi$ satisfies the equation of motion $\square \phi = 0$, the level-2 scalar descendant $\square \phi$ vanishes. Consequently, we expect only a single state at $\Delta = 5/2$ -- precisely as observed in the numerical result.
In contrast, for composite operators like $\phi^2$, $\square \phi^2 \neq 0$, and we indeed observe two scalar states near $\Delta = 3$, with the other being the primary $\phi^6$.
Similarly, two spin-2 states appear at $\Delta = 3$, corresponding to the stress-energy tensor and the spin-2 descendant of $\phi^2$, in full agreement with the expected operator content of the free scalar CFT.
These results strongly suggest that our model at the optimized parameter is well-approximated by the tricritical Ising point with negligible corrections from the $\phi^6$ interaction, i.e., the free scalar CFT.

\begin{figure}[t]
\centering
\includegraphics[width = 0.48\textwidth]{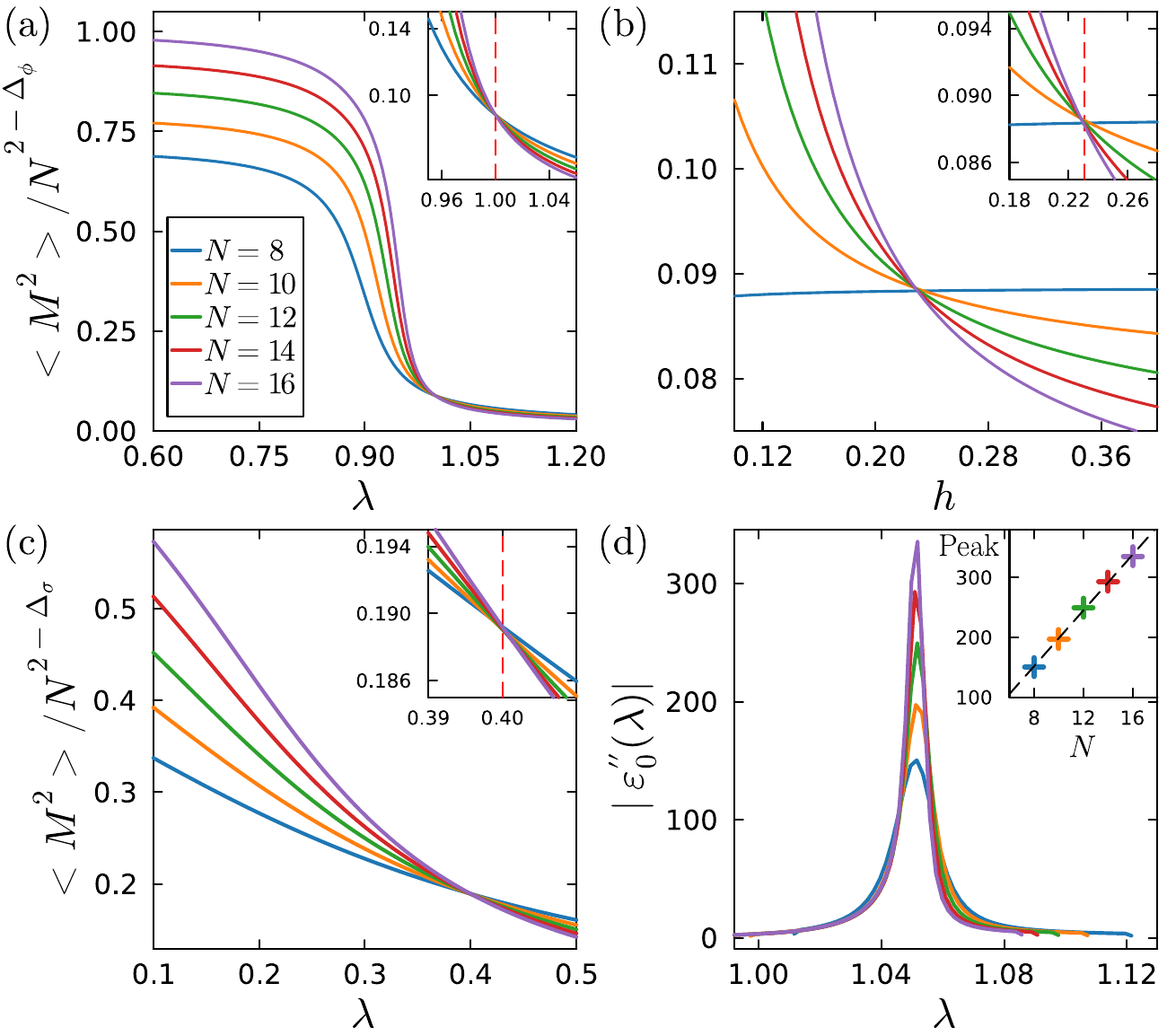} 
\caption{Finite-size scaling across different parts of the phase diagram. 
(a) Scaled magnetization, $\langle M^2\rangle/N^{2-\Delta_\phi}$ with $\Delta_\phi=1/2$, as $\lambda$ is varied through the tricritical point at fixed $h=0.23$. Inset magnifies the crossing around $\lambda=1$. (b) Same as (a) but fixing $\lambda=1$ and varying $h$. (c) Similar to (a) but crossing the 3D Ising line at a fixed $h=2$ by varying $\lambda$ and rescaling via $\Delta_\sigma=0.5181489$.  
(d) The second derivative of the ground state energy density, $\varepsilon_0$, with respect to $\lambda$ at fixed $h=0.05$. A scaling ansatz has been used to rescale the curves~\cite{SM} and the inset shows linear scaling of the peak value across the transition. The system sizes shown are color consistent in all the panels.
}
\label{fig:TCI FSS}
\end{figure}

To confirm the nature of Ising and tricritical Ising transitions, in Fig.~\ref{fig:TCI FSS} we perform detailed finite-size scaling analysis of the squared magnetization $\braket{M^2}$ in three ways: by taking a slice through the tricritical point along either horizontal ($\lambda$) direction [Fig.~\ref{fig:TCI FSS}(a)] or vertical ($h$) direction [Fig.~\ref{fig:TCI FSS}(b)], and by crossing the 3D Ising line at a fixed $h=2$ by varying $\lambda$ [Fig.~\ref{fig:TCI FSS}(c)]. In the first two cases, we rescale the data using $1/N^{2 - \Delta_\phi}$, with  $\Delta_\phi = 1/2$ corresponding to the Gaussian universality class, as logarithmic corrections are typically negligible at the system sizes accessible in numerics.
In the last case we use the conformal bootstrap value $\Delta_\sigma$ for the 3D Ising~\cite{Poland19}. We observe a clear crossing in Fig.~\ref{fig:TCI FSS}(a)-(b), consistent with $(\lambda, h) \approx (1, 0.23)$ being a tricritical point. Similarly, the data in Fig.~\ref{fig:TCI FSS}(c) is in agreement with 3D Ising universality class along the black dashed line in Fig.~\ref{fig:phase diagram}(a). 

To diagnose the first-order transition, we fix $h$ and examine the behavior of the ground state energy density $\varepsilon_0 = E_0/N$ as a function of $\lambda$. 
A first-order transition implies a kink in $\varepsilon_0(\lambda)$, typically manifesting as a sharp peak in its second derivative $\varepsilon_0''(\lambda) = d^2 \varepsilon_0(\lambda)/d\lambda^2$.
This is indeed observed in
\figref{fig:TCI FSS}(d). While the peak sharpens and its magnitude increases with system size, this growth appears to be linear [\figref{fig:TCI FSS}(d), inset], contrary to the naively expected exponential growth for a first-order transition.
This may be due to limited resolution of sampled points in $\lambda$ or due to the neighboring composite Fermi liquid phase at $h = 0$, $\lambda\approx 1$, potentially inducing a large correlation length and masking the exponential scaling.
Nevertheless, we emphasize that $\varepsilon_0''(\lambda)$ across the Ising transition line does not have any visible peak~\cite{SM}, pointing to a clear change in behavior around the tricritical point.

{\bf \em Free-boson algebra.---}Further evidence for the free scalar theory comes from the realization of the free boson algebra. In canonical quantization, the free scalar field $\phi$ on a sphere of radius $R$ has the following mode expansion~\cite{SM}:
\begin{equation}
\begin{gathered}
    \phi(\boldsymbol{\Omega}) = \frac{1}{\sqrt{R}} \sum_{\ell,m} \frac{1}{\sqrt{2\ell + 1}} \big( Y_{\ell,m}(\boldsymbol{\Omega}) \phi_{\ell,m} + \mathrm{h.c.} \big) \,,
\end{gathered}
\end{equation}
where $Y_{\ell,m}(\boldsymbol{\Omega})$ are the spherical harmonics, $\phi_{\ell,m}$ are annihilation operators corresponding to bosonic modes with angular momentum $(\ell, m)$ that satisfy $[\phi_{\ell,m}, \phi_{\ell',m'}^\dag] = \delta_{\ell,\ell'} \delta_{m,m'}$.
The full Hilbert space of the free scalar can thus be constructed as a Fock space of these modes. 
For instance, the scalar primaries $|\phi^n\rangle$ correspond to states with $n$ bosons in the $\ell = 0$ mode.
To probe this Fock space structure, we can study how pseudospin operators encode the boson creation and annihilation operators.
Let us write the total pseudospin operator $N_z = \int \big(n_{\uparrow}(\boldsymbol{\Omega}) - n_{\downarrow}(\boldsymbol{\Omega}) \big) d^2 \boldsymbol{\Omega}$ via the boson operators in the following symmetry-allowed form
\begin{equation}
\begin{aligned}
    \widetilde{N}_z = \frac{N_z}{R^{2 - \Delta_\phi}} = a_1 \phi_{0,0} + \frac{a_3}{R} \phi_{0,0}^3 + \mathrm{h.c.} + \ldots 
\end{aligned}
\end{equation}
where ellipsis are higher order terms, and we introduce a rescaling factor so that $a_1,a_3$ do not depend on the system size.
In a strict free scalar CFT, matrix elements of $N_z$ between Fock states satisfy
\begin{equation}
\label{eqn:free boson}
    p(n) = \langle \phi^n | \widetilde{N}_z^n | 0 \rangle = \sqrt{n!} \, a_1^n + \mathcal{O}(R^{-1})\,,
\end{equation}
where the subleading corrections account for mixing with higher-order terms like $\phi_0^3$.

\begin{figure}[t]
\centering
\includegraphics[width = 0.48\textwidth]{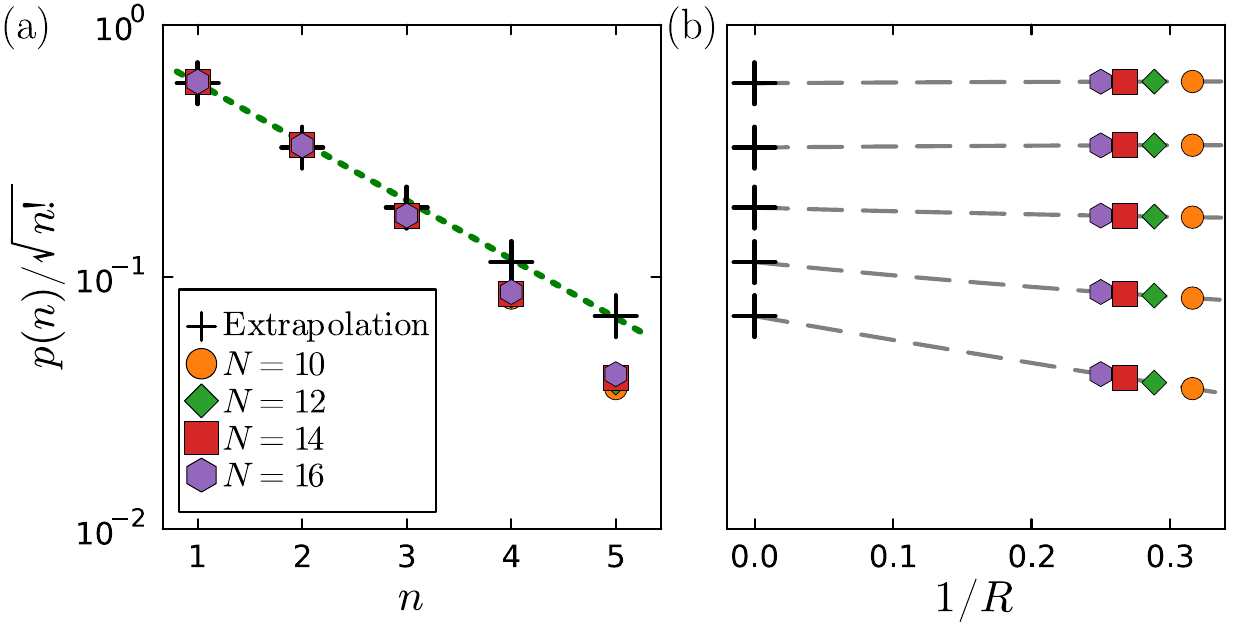}
\caption{Emergence of the free-boson algebra at the tricritical point. (a) $p(n)/\sqrt{n!}$ for  different system sizes. The extrapolated values (crosses) follow a linear dependence with $n$ on a log scale (dashed line), consistent with Eq.~(\ref{eqn:free boson}). (b) The extrapolated values shown in panel (a), obtained up to quadratic order in $1/R$. 
}
\label{fig:Tower}
\end{figure}

We numerically computed $p(n)$ and plot the result for $p(n)/\sqrt{n!}$ in \figref{fig:Tower}.
At a finite size, deviations do appear at large $n$.
After a finite-size scaling using \eqnref{eqn:free boson}, we obtain the values shown as crosses in \figref{fig:Tower}(a), which fall on a straight line—consistent with the prediction. By contrast, repeating the same calculation at a 3D Ising point, one can see a clear breakdown of Eq.~\eqref{eqn:free boson}~\cite{SM}. 
Similarly, the $\bbZ_2$-even total pseudospin operator takes the form $N_x = a_0 + a_2 \sum_{\ell,m} \phi_{\ell,m}^\dag \phi_{\ell,m} + \ldots$, which can be used to measure the boson number in each energy eigenstate. 
As shown in the SM~\cite{SM}, the measured expectation values in each eigenstate are indeed close to integers. This not only provides evidence for the realization of the free boson algebra, but also offers a practical diagnostic to distinguish certain primary states from descendants.

\begin{figure}[t]
\centering
\includegraphics[width = 0.48\textwidth]{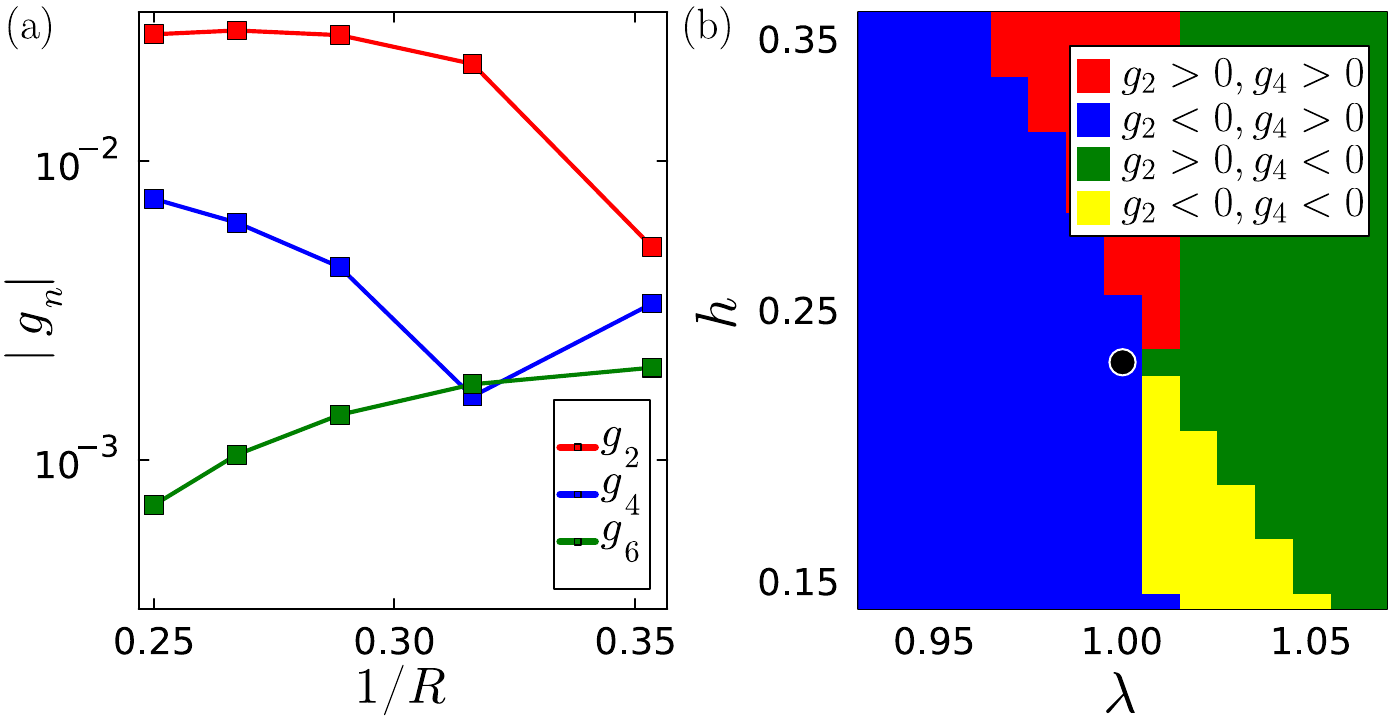}
\caption{Effective couplings near the tricritical point. We determine the effective coupling using six excited states that correspond to $\phi,\phi^2,\phi^3,\phi^4$ and $\partial\phi,\partial\phi^2$. The normalization is chosen such that the matrix elements of the operators $\phi^{2n}$ are independent of system size. (a) $g_2<0$, $g_6>0$ across all the system sizes, $g_4 < 0$ at $N=8$ and $g_4>0$ for $N>8$. (b) $g_6>0$ near the back dot and $g_6<0$ deeper inside the ferromagnetic phase (bulk of the blue region).}
\label{fig:CPT}
\end{figure}

{\bf \em Conformal perturbations.---}More rigorously, the system at the optimized parameter is a free scalar theory perturbed by small but finite interactions. We can quantitatively characterize it by the following effective description of the microscopic Hamiltonian~\cite{Lao:2023zis,Lauchli25}
\begin{equation}
\begin{aligned}
    H = \frac{v}{R} & H_{\text{CFT}} \\
    +&  \int \big( g_2 \, \phi^2(\boldsymbol{\Omega}) + g_4 \, \phi^4(\boldsymbol{\Omega}) + g_6 \, \phi^6(\boldsymbol{\Omega}) \big) d^2\boldsymbol{\Omega} \,,
\end{aligned}
	\label{eq:effective Hamiltonian}
\end{equation}
where $H_{\text{CFT}}$ is the free scalar CFT Hamiltonian, $v$ is a non-universal speed of light, and irrelevant perturbations are ignored. While the numerically obtained spectrum exhibits slight deviations from the exact CFT predictions, these discrepancies are well accounted for at linear order using \eqnref{eq:effective Hamiltonian}. \figref{fig:CPT}(a) shows the extracted effective couplings of the relevant terms are generally nonzero but sufficiently small, rendering the free scalar CFT an accurate description for moderate system sizes ($N \leq 16$).
In larger systems, we observe that the energy gap vanishes linearly with $1/\sqrt{N}$, and the two-point function continues to display the expected algebraic decay at long distances, as confirmed by density matrix renormalization group calculation up to $N{=}54$~\cite{SM}. 
These results suggest that the relevant couplings may remain negligible in the thermodynamic limit and the system at the optimized parameter still flows toward the free scalar.

One can also apply \eqnref{eq:effective Hamiltonian} to characterize the system in the vicinity of the optimized parameter. The results are shown in \figref{fig:CPT}(b). 
The Ising transition line aligns closely with the region where $g_4 > 0$ and $g_2$ changes sign, and the first-order transition line lies within the yellow regime, closer to the boundary where $g_2$ changes sign.
These results are broadly consistent with Ginzburg–Landau theory; the small discrepancies can be attributed to the positive coupling between $\phi^2$ and the background curvature. 
Most notably, the optimized point lies approximately at the intersection where both $g_2$ and $g_4$ change sign, providing additional evidence for its tricritical nature.

{\bf \em Conclusions.---}In this work, we used the Ising tricriticality to realize a free scalar CFT on the fuzzy sphere. The access to a free theory provides the minimal setting for understanding several aspects of the fuzzy sphere regularization.
While CFTs are characterized by the conformal algebra, charge-neutral excitations in the LLL are governed by the Girvin-MacDonald-Platzman (GMP) algebra~\cite{Girvin86}.
It is therefore natural to ask how the GMP algebra, when projected onto the low-energy subspace, encodes the conformal algebra~\cite{Cappelli:1992yv,Gromov17}.
Another question is how to extract all the universal CFT data only from the LLL since the higher Landau levels are conceptually part of the ultraviolet degrees of freedom within the fuzzy sphere regularization scheme.
For example, the free theory might have an efficient variational wave function description that allows us to understand the universal data in the orbital-cut entanglement and better extract the $F$-function~\cite{Li08,hu2025entropic}. 
Finally, an obvious next step is to explore how to realize free fermion CFT, which, together with the current result, could provide a proof of principle that the fuzzy sphere regularization can potentially encompass all renormalizable QFTs.

{\sl Note Added:} During the completion of this manuscript, we became aware of a related work~\cite{he2025freerealscalarcft}, which directly approaches the free scalar by drawing insights from its Euclidean action. 

\begin{acknowledgments}

{\bf \em Acknowledgments.---}We thank Ehud Altman, Zhehao Dai, Andrew Hallam, Yin-Chen He, Chong Wang, Michael Zaletel for helpful discussion. 
Computational portions of this research have made use of DiagHam~\cite{diagham} and FuzzifiED~\cite{FuzzifiED} software libraries, and they were carried out on ARC4 and AIRE, part of the High-Performance Computing facilities at the University of Leeds.
J.T., C.V. and Z.P. acknowledge support by the Leverhulme Trust Research Leadership Award RL-2019-015 and EPSRC Grant EP/Z533634/1. Statement of compliance with EPSRC policy framework on research data: This publication is theoretical work that does not require supporting research data. 
R.F. is supported by the Gordon and Betty Moore Foundation (Grant GBMF8688). 
This research was supported in part by grant NSF PHY-2309135 to the Kavli Institute for Theoretical Physics (KITP). Z.P. acknowledges support by the Erwin
Schrödinger International Institute for Mathematics and Physics.

\end{acknowledgments}

\bibliography{ref.bib}

\begin{thebibliography}{48}%
\makeatletter
\providecommand \@ifxundefined [1]{%
 \@ifx{#1\undefined}
}%
\providecommand \@ifnum [1]{%
 \ifnum #1\expandafter \@firstoftwo
 \else \expandafter \@secondoftwo
 \fi
}%
\providecommand \@ifx [1]{%
 \ifx #1\expandafter \@firstoftwo
 \else \expandafter \@secondoftwo
 \fi
}%
\providecommand \natexlab [1]{#1}%
\providecommand \enquote  [1]{``#1''}%
\providecommand \bibnamefont  [1]{#1}%
\providecommand \bibfnamefont [1]{#1}%
\providecommand \citenamefont [1]{#1}%
\providecommand \href@noop [0]{\@secondoftwo}%
\providecommand \href [0]{\begingroup \@sanitize@url \@href}%
\providecommand \@href[1]{\@@startlink{#1}\@@href}%
\providecommand \@@href[1]{\endgroup#1\@@endlink}%
\providecommand \@sanitize@url [0]{\catcode `\\12\catcode `\$12\catcode
  `\&12\catcode `\#12\catcode `\^12\catcode `\_12\catcode `\%12\relax}%
\providecommand \@@startlink[1]{}%
\providecommand \@@endlink[0]{}%
\providecommand \url  [0]{\begingroup\@sanitize@url \@url }%
\providecommand \@url [1]{\endgroup\@href {#1}{\urlprefix }}%
\providecommand \urlprefix  [0]{URL }%
\providecommand \Eprint [0]{\href }%
\providecommand \doibase [0]{https://doi.org/}%
\providecommand \selectlanguage [0]{\@gobble}%
\providecommand \bibinfo  [0]{\@secondoftwo}%
\providecommand \bibfield  [0]{\@secondoftwo}%
\providecommand \translation [1]{[#1]}%
\providecommand \BibitemOpen [0]{}%
\providecommand \bibitemStop [0]{}%
\providecommand \bibitemNoStop [0]{.\EOS\space}%
\providecommand \EOS [0]{\spacefactor3000\relax}%
\providecommand \BibitemShut  [1]{\csname bibitem#1\endcsname}%
\let\auto@bib@innerbib\@empty
\bibitem [{\citenamefont {Rychkov}(2016)}]{Rychkov:2016iqz}%
  \BibitemOpen
  \bibfield  {author} {\bibinfo {author} {\bibfnamefont {S.}~\bibnamefont
  {Rychkov}},\ }\href {https://doi.org/10.1007/978-3-319-43626-5} {\emph
  {\bibinfo {title} {{EPFL Lectures on Conformal Field Theory in
  D\ensuremath{>}= 3 Dimensions}}}},\ SpringerBriefs in Physics\ (\bibinfo
  {year} {2016})\ \Eprint {https://arxiv.org/abs/1601.05000} {arXiv:1601.05000
  [hep-th]} \BibitemShut {NoStop}%
\bibitem [{\citenamefont {Simmons-Duffin}(2017)}]{Simmons-Duffin:2016gjk}%
  \BibitemOpen
  \bibfield  {author} {\bibinfo {author} {\bibfnamefont {D.}~\bibnamefont
  {Simmons-Duffin}},\ }\bibfield  {title} {\bibinfo {title} {{The Conformal
  Bootstrap}},\ }in\ \href {https://doi.org/10.1142/9789813149441_0001} {\emph
  {\bibinfo {booktitle} {{Theoretical Advanced Study Institute in Elementary
  Particle Physics}: {New Frontiers in Fields and Strings}}}}\ (\bibinfo {year}
  {2017})\ pp.\ \bibinfo {pages} {1--74},\ \Eprint
  {https://arxiv.org/abs/1602.07982} {arXiv:1602.07982 [hep-th]} \BibitemShut
  {NoStop}%
\bibitem [{\citenamefont {Poland}\ \emph {et~al.}(2019)\citenamefont {Poland},
  \citenamefont {Rychkov},\ and\ \citenamefont {Vichi}}]{Poland19}%
  \BibitemOpen
  \bibfield  {author} {\bibinfo {author} {\bibfnamefont {D.}~\bibnamefont
  {Poland}}, \bibinfo {author} {\bibfnamefont {S.}~\bibnamefont {Rychkov}},\
  and\ \bibinfo {author} {\bibfnamefont {A.}~\bibnamefont {Vichi}},\ }\bibfield
   {title} {\bibinfo {title} {{The conformal bootstrap: Theory, numerical
  techniques, and applications}},\ }\href
  {https://doi.org/10.1103/RevModPhys.91.015002} {\bibfield  {journal}
  {\bibinfo  {journal} {Rev. Mod. Phys.}\ }\textbf {\bibinfo {volume} {91}},\
  \bibinfo {pages} {015002} (\bibinfo {year} {2019})}\BibitemShut {NoStop}%
\bibitem [{\citenamefont {Cardy}(1984)}]{Cardy84}%
  \BibitemOpen
  \bibfield  {author} {\bibinfo {author} {\bibfnamefont {J.~L.}\ \bibnamefont
  {Cardy}},\ }\bibfield  {title} {\bibinfo {title} {Conformal invariance and
  universality in finite-size scaling},\ }\href
  {https://doi.org/10.1088/0305-4470/17/7/003} {\bibfield  {journal} {\bibinfo
  {journal} {Journal of Physics A: Mathematical and General}\ }\textbf
  {\bibinfo {volume} {17}},\ \bibinfo {pages} {L385} (\bibinfo {year}
  {1984})}\BibitemShut {NoStop}%
\bibitem [{\citenamefont {Prange}\ and\ \citenamefont
  {Girvin}(1987)}]{Prange87}%
  \BibitemOpen
  \bibinfo {editor} {\bibfnamefont {R.~E.}\ \bibnamefont {Prange}}\ and\
  \bibinfo {editor} {\bibfnamefont {S.~M.}\ \bibnamefont {Girvin}},\ eds.,\
  \href@noop {} {\emph {\bibinfo {title} {The Quantum {Hall} Effect}}}\
  (\bibinfo  {publisher} {Springer-Verlag New York},\ \bibinfo {year}
  {1987})\BibitemShut {NoStop}%
\bibitem [{\citenamefont {Zhu}\ \emph {et~al.}(2023)\citenamefont {Zhu},
  \citenamefont {Han}, \citenamefont {Huffman}, \citenamefont {Hofmann},\ and\
  \citenamefont {He}}]{Zhu23}%
  \BibitemOpen
  \bibfield  {author} {\bibinfo {author} {\bibfnamefont {W.}~\bibnamefont
  {Zhu}}, \bibinfo {author} {\bibfnamefont {C.}~\bibnamefont {Han}}, \bibinfo
  {author} {\bibfnamefont {E.}~\bibnamefont {Huffman}}, \bibinfo {author}
  {\bibfnamefont {J.~S.}\ \bibnamefont {Hofmann}},\ and\ \bibinfo {author}
  {\bibfnamefont {Y.-C.}\ \bibnamefont {He}},\ }\bibfield  {title} {\bibinfo
  {title} {{Uncovering Conformal Symmetry in the 3D Ising Transition:
  State-Operator Correspondence from a Quantum Fuzzy Sphere Regularization}},\
  }\href {https://doi.org/10.1103/PhysRevX.13.021009} {\bibfield  {journal}
  {\bibinfo  {journal} {Phys. Rev. X}\ }\textbf {\bibinfo {volume} {13}},\
  \bibinfo {pages} {021009} (\bibinfo {year} {2023})}\BibitemShut {NoStop}%
\bibitem [{\citenamefont {Hu}\ \emph {et~al.}(2023{\natexlab{a}})\citenamefont
  {Hu}, \citenamefont {He},\ and\ \citenamefont {Zhu}}]{HuOPE2023}%
  \BibitemOpen
  \bibfield  {author} {\bibinfo {author} {\bibfnamefont {L.}~\bibnamefont
  {Hu}}, \bibinfo {author} {\bibfnamefont {Y.-C.}\ \bibnamefont {He}},\ and\
  \bibinfo {author} {\bibfnamefont {W.}~\bibnamefont {Zhu}},\ }\bibfield
  {title} {\bibinfo {title} {Operator product expansion coefficients of the 3d
  ising criticality via quantum fuzzy spheres},\ }\href
  {https://doi.org/10.1103/PhysRevLett.131.031601} {\bibfield  {journal}
  {\bibinfo  {journal} {Phys. Rev. Lett.}\ }\textbf {\bibinfo {volume} {131}},\
  \bibinfo {pages} {031601} (\bibinfo {year} {2023}{\natexlab{a}})}\BibitemShut
  {NoStop}%
\bibitem [{\citenamefont {Han}\ \emph {et~al.}(2023)\citenamefont {Han},
  \citenamefont {Hu}, \citenamefont {Zhu},\ and\ \citenamefont
  {He}}]{Han:2023yyb}%
  \BibitemOpen
  \bibfield  {author} {\bibinfo {author} {\bibfnamefont {C.}~\bibnamefont
  {Han}}, \bibinfo {author} {\bibfnamefont {L.}~\bibnamefont {Hu}}, \bibinfo
  {author} {\bibfnamefont {W.}~\bibnamefont {Zhu}},\ and\ \bibinfo {author}
  {\bibfnamefont {Y.-C.}\ \bibnamefont {He}},\ }\bibfield  {title} {\bibinfo
  {title} {{Conformal four-point correlators of the three-dimensional Ising
  transition via the quantum fuzzy sphere}},\ }\href
  {https://doi.org/10.1103/PhysRevB.108.235123} {\bibfield  {journal} {\bibinfo
   {journal} {Phys. Rev. B}\ }\textbf {\bibinfo {volume} {108}},\ \bibinfo
  {pages} {235123} (\bibinfo {year} {2023})},\ \Eprint
  {https://arxiv.org/abs/2306.04681} {arXiv:2306.04681 [cond-mat.stat-mech]}
  \BibitemShut {NoStop}%
\bibitem [{\citenamefont {Fardelli}\ \emph {et~al.}()\citenamefont {Fardelli},
  \citenamefont {Fitzpatrick},\ and\ \citenamefont
  {Katz}}]{fardelli2024constructinginfraredconformalgenerators}%
  \BibitemOpen
  \bibfield  {author} {\bibinfo {author} {\bibfnamefont {G.}~\bibnamefont
  {Fardelli}}, \bibinfo {author} {\bibfnamefont {A.~L.}\ \bibnamefont
  {Fitzpatrick}},\ and\ \bibinfo {author} {\bibfnamefont {E.}~\bibnamefont
  {Katz}},\ }\href {https://arxiv.org/abs/2409.02998} {\bibinfo {title}
  {Constructing the infrared conformal generators on the fuzzy sphere,
  arxiv:2409.02998}},\ \Eprint {https://arxiv.org/abs/2409.02998}
  {arXiv:2409.02998 [hep-th]} \BibitemShut {NoStop}%
\bibitem [{\citenamefont {Fan}()}]{fan2024noteexplicitconstructionconformal}%
  \BibitemOpen
  \bibfield  {author} {\bibinfo {author} {\bibfnamefont {R.}~\bibnamefont
  {Fan}},\ }\href {https://arxiv.org/abs/2409.08257} {\bibinfo {title} {Note on
  explicit construction of conformal generators on the fuzzy sphere,
  arxiv:2409.08257}},\ \Eprint {https://arxiv.org/abs/2409.08257}
  {arXiv:2409.08257 [hep-th]} \BibitemShut {NoStop}%
\bibitem [{\citenamefont {Hu}\ \emph {et~al.}(2025)\citenamefont {Hu},
  \citenamefont {Zhu},\ and\ \citenamefont {He}}]{hu2025entropic}%
  \BibitemOpen
  \bibfield  {author} {\bibinfo {author} {\bibfnamefont {L.}~\bibnamefont
  {Hu}}, \bibinfo {author} {\bibfnamefont {W.}~\bibnamefont {Zhu}},\ and\
  \bibinfo {author} {\bibfnamefont {Y.-C.}\ \bibnamefont {He}},\ }\bibfield
  {title} {\bibinfo {title} {Entropic f function of three-dimensional ising
  conformal field theory via fuzzy sphere regularization},\ }\href@noop {}
  {\bibfield  {journal} {\bibinfo  {journal} {Physical Review B}\ }\textbf
  {\bibinfo {volume} {111}},\ \bibinfo {pages} {155151} (\bibinfo {year}
  {2025})}\BibitemShut {NoStop}%
\bibitem [{\citenamefont {Hu}\ \emph {et~al.}(2024)\citenamefont {Hu},
  \citenamefont {He},\ and\ \citenamefont {Zhu}}]{Hu2024}%
  \BibitemOpen
  \bibfield  {author} {\bibinfo {author} {\bibfnamefont {L.}~\bibnamefont
  {Hu}}, \bibinfo {author} {\bibfnamefont {Y.-C.}\ \bibnamefont {He}},\ and\
  \bibinfo {author} {\bibfnamefont {W.}~\bibnamefont {Zhu}},\ }\bibfield
  {title} {\bibinfo {title} {Solving conformal defects in 3d conformal field
  theory using fuzzy sphere regularization},\ }\bibfield  {journal} {\bibinfo
  {journal} {Nature Communications}\ }\textbf {\bibinfo {volume} {15}},\ \href
  {https://doi.org/10.1038/s41467-024-47978-y} {10.1038/s41467-024-47978-y}
  (\bibinfo {year} {2024})\BibitemShut {NoStop}%
\bibitem [{\citenamefont {Zhou}\ \emph
  {et~al.}(2024{\natexlab{a}})\citenamefont {Zhou}, \citenamefont {Gaiotto},
  \citenamefont {He},\ and\ \citenamefont {Zou}}]{Zhou2024}%
  \BibitemOpen
  \bibfield  {author} {\bibinfo {author} {\bibfnamefont {Z.}~\bibnamefont
  {Zhou}}, \bibinfo {author} {\bibfnamefont {D.}~\bibnamefont {Gaiotto}},
  \bibinfo {author} {\bibfnamefont {Y.-C.}\ \bibnamefont {He}},\ and\ \bibinfo
  {author} {\bibfnamefont {Y.}~\bibnamefont {Zou}},\ }\bibfield  {title}
  {\bibinfo {title} {The $g$-function and defect changing operators from
  wavefunction overlap on a fuzzy sphere},\ }\bibfield  {journal} {\bibinfo
  {journal} {SciPost Physics}\ }\textbf {\bibinfo {volume} {17}},\ \href
  {https://doi.org/10.21468/scipostphys.17.1.021}
  {10.21468/scipostphys.17.1.021} (\bibinfo {year}
  {2024}{\natexlab{a}})\BibitemShut {NoStop}%
\bibitem [{\citenamefont {Zhou}\ and\ \citenamefont {Zou}(2025)}]{Zhou2025}%
  \BibitemOpen
  \bibfield  {author} {\bibinfo {author} {\bibfnamefont {Z.}~\bibnamefont
  {Zhou}}\ and\ \bibinfo {author} {\bibfnamefont {Y.}~\bibnamefont {Zou}},\
  }\bibfield  {title} {\bibinfo {title} {Studying the 3d ising surface cfts on
  the fuzzy sphere},\ }\bibfield  {journal} {\bibinfo  {journal} {SciPost
  Physics}\ }\textbf {\bibinfo {volume} {18}},\ \href
  {https://doi.org/10.21468/scipostphys.18.1.031}
  {10.21468/scipostphys.18.1.031} (\bibinfo {year} {2025})\BibitemShut
  {NoStop}%
\bibitem [{\citenamefont
  {Dedushenko}(2024)}]{dedushenko2024isingbcftfuzzyhemisphere}%
  \BibitemOpen
  \bibfield  {author} {\bibinfo {author} {\bibfnamefont {M.}~\bibnamefont
  {Dedushenko}},\ }\href {https://arxiv.org/abs/2407.15948} {\bibinfo {title}
  {Ising bcft from fuzzy hemisphere}} (\bibinfo {year} {2024}),\ \Eprint
  {https://arxiv.org/abs/2407.15948} {arXiv:2407.15948 [hep-th]} \BibitemShut
  {NoStop}%
\bibitem [{\citenamefont {Han}\ \emph {et~al.}(2024)\citenamefont {Han},
  \citenamefont {Hu},\ and\ \citenamefont {Zhu}}]{Han24}%
  \BibitemOpen
  \bibfield  {author} {\bibinfo {author} {\bibfnamefont {C.}~\bibnamefont
  {Han}}, \bibinfo {author} {\bibfnamefont {L.}~\bibnamefont {Hu}},\ and\
  \bibinfo {author} {\bibfnamefont {W.}~\bibnamefont {Zhu}},\ }\bibfield
  {title} {\bibinfo {title} {{Conformal operator content of the Wilson-Fisher
  transition on fuzzy sphere bilayers}},\ }\href
  {https://doi.org/10.1103/PhysRevB.110.115113} {\bibfield  {journal} {\bibinfo
   {journal} {Phys. Rev. B}\ }\textbf {\bibinfo {volume} {110}},\ \bibinfo
  {pages} {115113} (\bibinfo {year} {2024})}\BibitemShut {NoStop}%
\bibitem [{\citenamefont {Voinea}\ \emph {et~al.}(2024)\citenamefont {Voinea},
  \citenamefont {Fan}, \citenamefont {Regnault},\ and\ \citenamefont
  {Papić}}]{voinea2024regularizing}%
  \BibitemOpen
  \bibfield  {author} {\bibinfo {author} {\bibfnamefont {C.}~\bibnamefont
  {Voinea}}, \bibinfo {author} {\bibfnamefont {R.}~\bibnamefont {Fan}},
  \bibinfo {author} {\bibfnamefont {N.}~\bibnamefont {Regnault}},\ and\
  \bibinfo {author} {\bibfnamefont {Z.}~\bibnamefont {Papić}},\ }\href
  {https://arxiv.org/abs/2411.15299} {\bibinfo {title} {Regularizing 3d
  conformal field theories via anyons on the fuzzy sphere}} (\bibinfo {year}
  {2024}),\ \Eprint {https://arxiv.org/abs/2411.15299} {arXiv:2411.15299
  [cond-mat.stat-mech]} \BibitemShut {NoStop}%
\bibitem [{\citenamefont {Läuchli}\ \emph {et~al.}(2025)\citenamefont
  {Läuchli}, \citenamefont {Herviou}, \citenamefont {Wilhelm},\ and\
  \citenamefont {Rychkov}}]{Lauchli25}%
  \BibitemOpen
  \bibfield  {author} {\bibinfo {author} {\bibfnamefont {A.~M.}\ \bibnamefont
  {Läuchli}}, \bibinfo {author} {\bibfnamefont {L.}~\bibnamefont {Herviou}},
  \bibinfo {author} {\bibfnamefont {P.~H.}\ \bibnamefont {Wilhelm}},\ and\
  \bibinfo {author} {\bibfnamefont {S.}~\bibnamefont {Rychkov}},\ }\href
  {https://arxiv.org/abs/2504.00842} {\bibinfo {title} {Exact diagonalization,
  matrix product states and conformal perturbation theory study of a 3d ising
  fuzzy sphere model}} (\bibinfo {year} {2025}),\ \Eprint
  {https://arxiv.org/abs/2504.00842} {arXiv:2504.00842 [cond-mat.stat-mech]}
  \BibitemShut {NoStop}%
\bibitem [{\citenamefont {Ippoliti}\ \emph {et~al.}(2018)\citenamefont
  {Ippoliti}, \citenamefont {Mong}, \citenamefont {Assaad},\ and\ \citenamefont
  {Zaletel}}]{Ippoliti:2018prb}%
  \BibitemOpen
  \bibfield  {author} {\bibinfo {author} {\bibfnamefont {M.}~\bibnamefont
  {Ippoliti}}, \bibinfo {author} {\bibfnamefont {R.~S.~K.}\ \bibnamefont
  {Mong}}, \bibinfo {author} {\bibfnamefont {F.~F.}\ \bibnamefont {Assaad}},\
  and\ \bibinfo {author} {\bibfnamefont {M.~P.}\ \bibnamefont {Zaletel}},\
  }\bibfield  {title} {\bibinfo {title} {Half-filled landau levels: A continuum
  and sign-free regularization for three-dimensional quantum critical points},\
  }\href {https://doi.org/10.1103/PhysRevB.98.235108} {\bibfield  {journal}
  {\bibinfo  {journal} {Phys. Rev. B}\ }\textbf {\bibinfo {volume} {98}},\
  \bibinfo {pages} {235108} (\bibinfo {year} {2018})}\BibitemShut {NoStop}%
\bibitem [{\citenamefont {Zhou}\ \emph
  {et~al.}(2024{\natexlab{b}})\citenamefont {Zhou}, \citenamefont {Hu},
  \citenamefont {Zhu},\ and\ \citenamefont {He}}]{Zhou:2024qfi}%
  \BibitemOpen
  \bibfield  {author} {\bibinfo {author} {\bibfnamefont {Z.}~\bibnamefont
  {Zhou}}, \bibinfo {author} {\bibfnamefont {L.}~\bibnamefont {Hu}}, \bibinfo
  {author} {\bibfnamefont {W.}~\bibnamefont {Zhu}},\ and\ \bibinfo {author}
  {\bibfnamefont {Y.-C.}\ \bibnamefont {He}},\ }\bibfield  {title} {\bibinfo
  {title} {So(5) deconfined phase transition under the fuzzy-sphere microscope:
  Approximate conformal symmetry, pseudo-criticality, and operator spectrum},\
  }\href {https://doi.org/10.1103/PhysRevX.14.021044} {\bibfield  {journal}
  {\bibinfo  {journal} {Phys. Rev. X}\ }\textbf {\bibinfo {volume} {14}},\
  \bibinfo {pages} {021044} (\bibinfo {year} {2024}{\natexlab{b}})}\BibitemShut
  {NoStop}%
\bibitem [{\citenamefont {Chen}\ \emph
  {et~al.}(2024{\natexlab{a}})\citenamefont {Chen}, \citenamefont {Zhang},
  \citenamefont {Wang}, \citenamefont {Sun},\ and\ \citenamefont
  {Meng}}]{chen24a}%
  \BibitemOpen
  \bibfield  {author} {\bibinfo {author} {\bibfnamefont {B.-B.}\ \bibnamefont
  {Chen}}, \bibinfo {author} {\bibfnamefont {X.}~\bibnamefont {Zhang}},
  \bibinfo {author} {\bibfnamefont {Y.}~\bibnamefont {Wang}}, \bibinfo {author}
  {\bibfnamefont {K.}~\bibnamefont {Sun}},\ and\ \bibinfo {author}
  {\bibfnamefont {Z.~Y.}\ \bibnamefont {Meng}},\ }\bibfield  {title} {\bibinfo
  {title} {Phases of $(2+1)\mathrm{D}$ so(5) nonlinear sigma model with a
  topological term on a sphere: Multicritical point and disorder phase},\
  }\href {https://doi.org/10.1103/PhysRevLett.132.246503} {\bibfield  {journal}
  {\bibinfo  {journal} {Phys. Rev. Lett.}\ }\textbf {\bibinfo {volume} {132}},\
  \bibinfo {pages} {246503} (\bibinfo {year} {2024}{\natexlab{a}})}\BibitemShut
  {NoStop}%
\bibitem [{\citenamefont {Chen}\ \emph
  {et~al.}(2024{\natexlab{b}})\citenamefont {Chen}, \citenamefont {Zhang},\
  and\ \citenamefont {Meng}}]{chen24b}%
  \BibitemOpen
  \bibfield  {author} {\bibinfo {author} {\bibfnamefont {B.-B.}\ \bibnamefont
  {Chen}}, \bibinfo {author} {\bibfnamefont {X.}~\bibnamefont {Zhang}},\ and\
  \bibinfo {author} {\bibfnamefont {Z.~Y.}\ \bibnamefont {Meng}},\ }\bibfield
  {title} {\bibinfo {title} {Emergent conformal symmetry at the multicritical
  point of $(2+1)\mathrm{D}$ so(5) model with wess-zumino-witten term on a
  sphere},\ }\href {https://doi.org/10.1103/PhysRevB.110.125153} {\bibfield
  {journal} {\bibinfo  {journal} {Phys. Rev. B}\ }\textbf {\bibinfo {volume}
  {110}},\ \bibinfo {pages} {125153} (\bibinfo {year}
  {2024}{\natexlab{b}})}\BibitemShut {NoStop}%
\bibitem [{\citenamefont {Yang}\ \emph {et~al.}(2025)\citenamefont {Yang},
  \citenamefont {Yue}, \citenamefont {Tang}, \citenamefont {Han}, \citenamefont
  {Zhu},\ and\ \citenamefont {Chen}}]{yang2025microscopic}%
  \BibitemOpen
  \bibfield  {author} {\bibinfo {author} {\bibfnamefont {S.}~\bibnamefont
  {Yang}}, \bibinfo {author} {\bibfnamefont {Y.-G.}\ \bibnamefont {Yue}},
  \bibinfo {author} {\bibfnamefont {Y.}~\bibnamefont {Tang}}, \bibinfo {author}
  {\bibfnamefont {C.}~\bibnamefont {Han}}, \bibinfo {author} {\bibfnamefont
  {W.}~\bibnamefont {Zhu}},\ and\ \bibinfo {author} {\bibfnamefont
  {Y.}~\bibnamefont {Chen}},\ }\href {https://arxiv.org/abs/2501.14320}
  {\bibinfo {title} {Microscopic study of 3d potts phase transition via fuzzy
  sphere regularization}} (\bibinfo {year} {2025}),\ \Eprint
  {https://arxiv.org/abs/2501.14320} {arXiv:2501.14320 [cond-mat.stat-mech]}
  \BibitemShut {NoStop}%
\bibitem [{\citenamefont {Fan}\ \emph {et~al.}(2025)\citenamefont {Fan},
  \citenamefont {Dong},\ and\ \citenamefont
  {Vishwanath}}]{fan2025simulatingnonunitaryyangleeconformal}%
  \BibitemOpen
  \bibfield  {author} {\bibinfo {author} {\bibfnamefont {R.}~\bibnamefont
  {Fan}}, \bibinfo {author} {\bibfnamefont {J.}~\bibnamefont {Dong}},\ and\
  \bibinfo {author} {\bibfnamefont {A.}~\bibnamefont {Vishwanath}},\ }\href
  {https://arxiv.org/abs/2505.06342} {\bibinfo {title} {Simulating the
  non-unitary yang-lee conformal field theory on the fuzzy sphere}} (\bibinfo
  {year} {2025}),\ \Eprint {https://arxiv.org/abs/2505.06342} {arXiv:2505.06342
  [cond-mat.str-el]} \BibitemShut {NoStop}%
\bibitem [{\citenamefont {Cruz}\ \emph {et~al.}(2025)\citenamefont {Cruz},
  \citenamefont {Klebanov}, \citenamefont {Tarnopolsky},\ and\ \citenamefont
  {Xin}}]{cruz2025yangleequantumcriticalityvarious}%
  \BibitemOpen
  \bibfield  {author} {\bibinfo {author} {\bibfnamefont {E.~A.}\ \bibnamefont
  {Cruz}}, \bibinfo {author} {\bibfnamefont {I.~R.}\ \bibnamefont {Klebanov}},
  \bibinfo {author} {\bibfnamefont {G.}~\bibnamefont {Tarnopolsky}},\ and\
  \bibinfo {author} {\bibfnamefont {Y.}~\bibnamefont {Xin}},\ }\href
  {https://arxiv.org/abs/2505.06369} {\bibinfo {title} {Yang-lee quantum
  criticality in various dimensions}} (\bibinfo {year} {2025}),\ \Eprint
  {https://arxiv.org/abs/2505.06369} {arXiv:2505.06369 [hep-th]} \BibitemShut
  {NoStop}%
\bibitem [{\citenamefont {Miro}\ and\ \citenamefont
  {Delouche}(2025)}]{miro2025flowingisingmodelfuzzy}%
  \BibitemOpen
  \bibfield  {author} {\bibinfo {author} {\bibfnamefont {J.~E.}\ \bibnamefont
  {Miro}}\ and\ \bibinfo {author} {\bibfnamefont {O.}~\bibnamefont
  {Delouche}},\ }\href {https://arxiv.org/abs/2505.07655} {\bibinfo {title}
  {Flowing from the ising model on the fuzzy sphere to the 3d lee-yang cft}}
  (\bibinfo {year} {2025}),\ \Eprint {https://arxiv.org/abs/2505.07655}
  {arXiv:2505.07655 [hep-th]} \BibitemShut {NoStop}%
\bibitem [{\citenamefont {Zhou}\ and\ \citenamefont {He}(2024)}]{Zhou24b}%
  \BibitemOpen
  \bibfield  {author} {\bibinfo {author} {\bibfnamefont {Z.}~\bibnamefont
  {Zhou}}\ and\ \bibinfo {author} {\bibfnamefont {Y.-C.}\ \bibnamefont {He}},\
  }\href {https://arxiv.org/abs/2410.00087} {\bibinfo {title} {{A new series of
  3D CFTs with $\mathrm{Sp}(N)$ global symmetry on fuzzy sphere}}} (\bibinfo
  {year} {2024}),\ \Eprint {https://arxiv.org/abs/2410.00087} {arXiv:2410.00087
  [hep-th]} \BibitemShut {NoStop}%
\bibitem [{\citenamefont {Cardy}(1996)}]{cardy1996scaling}%
  \BibitemOpen
  \bibfield  {author} {\bibinfo {author} {\bibfnamefont {J.}~\bibnamefont
  {Cardy}},\ }\href@noop {} {\emph {\bibinfo {title} {Scaling and
  renormalization in statistical physics}}},\ Vol.~\bibinfo {volume} {5}\
  (\bibinfo  {publisher} {Cambridge university press},\ \bibinfo {year}
  {1996})\BibitemShut {NoStop}%
\bibitem [{\citenamefont
  {Henriksson}(2025)}]{henriksson2025tricriticalisingcftconformal}%
  \BibitemOpen
  \bibfield  {author} {\bibinfo {author} {\bibfnamefont {J.}~\bibnamefont
  {Henriksson}},\ }\href {https://arxiv.org/abs/2501.18711} {\bibinfo {title}
  {The tricritical ising cft and conformal bootstrap}} (\bibinfo {year}
  {2025}),\ \Eprint {https://arxiv.org/abs/2501.18711} {arXiv:2501.18711
  [hep-th]} \BibitemShut {NoStop}%
\bibitem [{\citenamefont {Blume}(1966)}]{Blume1966}%
  \BibitemOpen
  \bibfield  {author} {\bibinfo {author} {\bibfnamefont {M.}~\bibnamefont
  {Blume}},\ }\bibfield  {title} {\bibinfo {title} {Theory of the first-order
  magnetic phase change in u${\mathrm{o}}_{2}$},\ }\href
  {https://doi.org/10.1103/PhysRev.141.517} {\bibfield  {journal} {\bibinfo
  {journal} {Phys. Rev.}\ }\textbf {\bibinfo {volume} {141}},\ \bibinfo {pages}
  {517} (\bibinfo {year} {1966})}\BibitemShut {NoStop}%
\bibitem [{\citenamefont {Capel}(1966)}]{Capel1966}%
  \BibitemOpen
  \bibfield  {author} {\bibinfo {author} {\bibfnamefont {H.}~\bibnamefont
  {Capel}},\ }\bibfield  {title} {\bibinfo {title} {On the possibility of
  first-order phase transitions in ising systems of triplet ions with
  zero-field splitting},\ }\href
  {https://doi.org/https://doi.org/10.1016/0031-8914(66)90027-9} {\bibfield
  {journal} {\bibinfo  {journal} {Physica}\ }\textbf {\bibinfo {volume} {32}},\
  \bibinfo {pages} {966} (\bibinfo {year} {1966})}\BibitemShut {NoStop}%
\bibitem [{\citenamefont {Haldane}(1983)}]{Haldane:1983xm}%
  \BibitemOpen
  \bibfield  {author} {\bibinfo {author} {\bibfnamefont {F.~D.~M.}\
  \bibnamefont {Haldane}},\ }\bibfield  {title} {\bibinfo {title} {{Fractional
  quantization of the Hall effect: A Hierarchy of incompressible quantum fluid
  states}},\ }\href {https://doi.org/10.1103/PhysRevLett.51.605} {\bibfield
  {journal} {\bibinfo  {journal} {Phys. Rev. Lett.}\ }\textbf {\bibinfo
  {volume} {51}},\ \bibinfo {pages} {605} (\bibinfo {year} {1983})}\BibitemShut
  {NoStop}%
\bibitem [{SM()}]{SM}%
  \BibitemOpen
  \href@noop {} {}\bibinfo {note} {See the Supplemental Online Material for
  details of the derivations and further numerical results.}\BibitemShut
  {Stop}%
\bibitem [{\citenamefont {Grover}\ \emph {et~al.}(2014)\citenamefont {Grover},
  \citenamefont {Sheng},\ and\ \citenamefont {Vishwanath}}]{Grover:2013rc}%
  \BibitemOpen
  \bibfield  {author} {\bibinfo {author} {\bibfnamefont {T.}~\bibnamefont
  {Grover}}, \bibinfo {author} {\bibfnamefont {D.~N.}\ \bibnamefont {Sheng}},\
  and\ \bibinfo {author} {\bibfnamefont {A.}~\bibnamefont {Vishwanath}},\
  }\bibfield  {title} {\bibinfo {title} {Emergent space-time supersymmetry at
  the boundary of a topological phase},\ }\href
  {https://doi.org/10.1126/science.1248253} {\bibfield  {journal} {\bibinfo
  {journal} {Science}\ }\textbf {\bibinfo {volume} {344}},\ \bibinfo {pages}
  {280} (\bibinfo {year} {2014})}\BibitemShut {NoStop}%
\bibitem [{\citenamefont {Rahmani}\ \emph {et~al.}(2015)\citenamefont
  {Rahmani}, \citenamefont {Zhu}, \citenamefont {Franz},\ and\ \citenamefont
  {Affleck}}]{Rahmani:2015}%
  \BibitemOpen
  \bibfield  {author} {\bibinfo {author} {\bibfnamefont {A.}~\bibnamefont
  {Rahmani}}, \bibinfo {author} {\bibfnamefont {X.}~\bibnamefont {Zhu}},
  \bibinfo {author} {\bibfnamefont {M.}~\bibnamefont {Franz}},\ and\ \bibinfo
  {author} {\bibfnamefont {I.}~\bibnamefont {Affleck}},\ }\bibfield  {title}
  {\bibinfo {title} {Emergent supersymmetry from strongly interacting majorana
  zero modes},\ }\href {https://doi.org/10.1103/PhysRevLett.115.166401}
  {\bibfield  {journal} {\bibinfo  {journal} {Phys. Rev. Lett.}\ }\textbf
  {\bibinfo {volume} {115}},\ \bibinfo {pages} {166401} (\bibinfo {year}
  {2015})}\BibitemShut {NoStop}%
\bibitem [{\citenamefont {O'Brien}\ and\ \citenamefont
  {Fendley}(2018)}]{OBrien:2017wmx}%
  \BibitemOpen
  \bibfield  {author} {\bibinfo {author} {\bibfnamefont {E.}~\bibnamefont
  {O'Brien}}\ and\ \bibinfo {author} {\bibfnamefont {P.}~\bibnamefont
  {Fendley}},\ }\bibfield  {title} {\bibinfo {title} {Lattice supersymmetry and
  order-disorder coexistence in the tricritical ising model},\ }\href
  {https://doi.org/10.1103/PhysRevLett.120.206403} {\bibfield  {journal}
  {\bibinfo  {journal} {Phys. Rev. Lett.}\ }\textbf {\bibinfo {volume} {120}},\
  \bibinfo {pages} {206403} (\bibinfo {year} {2018})}\BibitemShut {NoStop}%
\bibitem [{\citenamefont {Halperin}\ \emph {et~al.}(1993)\citenamefont
  {Halperin}, \citenamefont {Lee},\ and\ \citenamefont {Read}}]{Halperin93}%
  \BibitemOpen
  \bibfield  {author} {\bibinfo {author} {\bibfnamefont {B.~I.}\ \bibnamefont
  {Halperin}}, \bibinfo {author} {\bibfnamefont {P.~A.}\ \bibnamefont {Lee}},\
  and\ \bibinfo {author} {\bibfnamefont {N.}~\bibnamefont {Read}},\ }\bibfield
  {title} {\bibinfo {title} {Theory of the half-filled {Landau} level},\ }\href
  {https://doi.org/10.1103/PhysRevB.47.7312} {\bibfield  {journal} {\bibinfo
  {journal} {Phys. Rev. B}\ }\textbf {\bibinfo {volume} {47}},\ \bibinfo
  {pages} {7312} (\bibinfo {year} {1993})}\BibitemShut {NoStop}%
\bibitem [{\citenamefont {Lao}\ and\ \citenamefont
  {Rychkov}(2023)}]{Lao:2023zis}%
  \BibitemOpen
  \bibfield  {author} {\bibinfo {author} {\bibfnamefont {B.-X.}\ \bibnamefont
  {Lao}}\ and\ \bibinfo {author} {\bibfnamefont {S.}~\bibnamefont {Rychkov}},\
  }\bibfield  {title} {\bibinfo {title} {{3D Ising CFT and exact
  diagonalization on icosahedron: The power of conformal perturbation
  theory}},\ }\href {https://doi.org/10.21468/SciPostPhys.15.6.243} {\bibfield
  {journal} {\bibinfo  {journal} {SciPost Phys.}\ }\textbf {\bibinfo {volume}
  {15}},\ \bibinfo {pages} {243} (\bibinfo {year} {2023})},\ \Eprint
  {https://arxiv.org/abs/2307.02540} {arXiv:2307.02540 [hep-th]} \BibitemShut
  {NoStop}%
\bibitem [{\citenamefont {Girvin}\ \emph {et~al.}(1986)\citenamefont {Girvin},
  \citenamefont {MacDonald},\ and\ \citenamefont {Platzman}}]{Girvin86}%
  \BibitemOpen
  \bibfield  {author} {\bibinfo {author} {\bibfnamefont {S.~M.}\ \bibnamefont
  {Girvin}}, \bibinfo {author} {\bibfnamefont {A.~H.}\ \bibnamefont
  {MacDonald}},\ and\ \bibinfo {author} {\bibfnamefont {P.~M.}\ \bibnamefont
  {Platzman}},\ }\bibfield  {title} {\bibinfo {title} {Magneto-roton theory of
  collective excitations in the fractional quantum {Hall} effect},\ }\href
  {https://doi.org/10.1103/PhysRevB.33.2481} {\bibfield  {journal} {\bibinfo
  {journal} {Phys. Rev. B}\ }\textbf {\bibinfo {volume} {33}},\ \bibinfo
  {pages} {2481} (\bibinfo {year} {1986})}\BibitemShut {NoStop}%
\bibitem [{\citenamefont {Cappelli}\ \emph {et~al.}(1993)\citenamefont
  {Cappelli}, \citenamefont {Trugenberger},\ and\ \citenamefont
  {Zemba}}]{Cappelli:1992yv}%
  \BibitemOpen
  \bibfield  {author} {\bibinfo {author} {\bibfnamefont {A.}~\bibnamefont
  {Cappelli}}, \bibinfo {author} {\bibfnamefont {C.~A.}\ \bibnamefont
  {Trugenberger}},\ and\ \bibinfo {author} {\bibfnamefont {G.~R.}\ \bibnamefont
  {Zemba}},\ }\bibfield  {title} {\bibinfo {title} {{Infinite symmetry in the
  quantum Hall effect}},\ }\href {https://doi.org/10.1016/0550-3213(93)90660-H}
  {\bibfield  {journal} {\bibinfo  {journal} {Nucl. Phys. B}\ }\textbf
  {\bibinfo {volume} {396}},\ \bibinfo {pages} {465} (\bibinfo {year}
  {1993})},\ \Eprint {https://arxiv.org/abs/hep-th/9206027}
  {arXiv:hep-th/9206027} \BibitemShut {NoStop}%
\bibitem [{\citenamefont {Gromov}\ and\ \citenamefont {Son}(2017)}]{Gromov17}%
  \BibitemOpen
  \bibfield  {author} {\bibinfo {author} {\bibfnamefont {A.}~\bibnamefont
  {Gromov}}\ and\ \bibinfo {author} {\bibfnamefont {D.~T.}\ \bibnamefont
  {Son}},\ }\bibfield  {title} {\bibinfo {title} {Bimetric theory of fractional
  quantum {Hall} states},\ }\href {https://doi.org/10.1103/PhysRevX.7.041032}
  {\bibfield  {journal} {\bibinfo  {journal} {Phys. Rev. X}\ }\textbf {\bibinfo
  {volume} {7}},\ \bibinfo {pages} {041032} (\bibinfo {year}
  {2017})}\BibitemShut {NoStop}%
\bibitem [{\citenamefont {Li}\ and\ \citenamefont {Haldane}(2008)}]{Li08}%
  \BibitemOpen
  \bibfield  {author} {\bibinfo {author} {\bibfnamefont {H.}~\bibnamefont
  {Li}}\ and\ \bibinfo {author} {\bibfnamefont {F.~D.~M.}\ \bibnamefont
  {Haldane}},\ }\bibfield  {title} {\bibinfo {title} {Entanglement spectrum as
  a generalization of entanglement entropy: Identification of topological order
  in non-abelian fractional quantum {Hall} effect states},\ }\href
  {https://doi.org/10.1103/PhysRevLett.101.010504} {\bibfield  {journal}
  {\bibinfo  {journal} {Phys. Rev. Lett.}\ }\textbf {\bibinfo {volume} {101}},\
  \bibinfo {pages} {010504} (\bibinfo {year} {2008})}\BibitemShut {NoStop}%
\bibitem [{\citenamefont {He}(2025)}]{he2025freerealscalarcft}%
  \BibitemOpen
  \bibfield  {author} {\bibinfo {author} {\bibfnamefont {Y.-C.}\ \bibnamefont
  {He}},\ }\href {https://arxiv.org/abs/2506.14904} {\bibinfo {title} {Free
  real scalar cft on fuzzy sphere: spectrum, algebra and wavefunction ansatz}}
  (\bibinfo {year} {2025}),\ \Eprint {https://arxiv.org/abs/2506.14904}
  {arXiv:2506.14904 [hep-th]} \BibitemShut {NoStop}%
\bibitem [{dia()}]{diagham}%
  \BibitemOpen
  \href@noop {} {}\bibinfo {note} {Diag{H}am,
  \url{https://www.nick-ux.org/diagham}}\BibitemShut {NoStop}%
\bibitem [{\citenamefont {Zhou}(2025)}]{FuzzifiED}%
  \BibitemOpen
  \bibfield  {author} {\bibinfo {author} {\bibfnamefont {Z.}~\bibnamefont
  {Zhou}},\ }\href {https://arxiv.org/abs/2503.00100} {\bibinfo {title}
  {{FuzzifiED -- Julia package for numerics on the fuzzy sphere}}} (\bibinfo
  {year} {2025}),\ \Eprint {https://arxiv.org/abs/2503.00100} {arXiv:2503.00100
  [cond-mat.str-el]} \BibitemShut {NoStop}%
\bibitem [{\citenamefont {Moon}\ \emph {et~al.}(1995)\citenamefont {Moon},
  \citenamefont {Mori}, \citenamefont {Yang}, \citenamefont {Girvin},
  \citenamefont {MacDonald}, \citenamefont {Zheng}, \citenamefont {Yoshioka},\
  and\ \citenamefont {Zhang}}]{moon199513SpontaneousInterlayer}%
  \BibitemOpen
  \bibfield  {author} {\bibinfo {author} {\bibfnamefont {K.}~\bibnamefont
  {Moon}}, \bibinfo {author} {\bibfnamefont {H.}~\bibnamefont {Mori}}, \bibinfo
  {author} {\bibfnamefont {K.}~\bibnamefont {Yang}}, \bibinfo {author}
  {\bibfnamefont {S.~M.}\ \bibnamefont {Girvin}}, \bibinfo {author}
  {\bibfnamefont {A.~H.}\ \bibnamefont {MacDonald}}, \bibinfo {author}
  {\bibfnamefont {L.}~\bibnamefont {Zheng}}, \bibinfo {author} {\bibfnamefont
  {D.}~\bibnamefont {Yoshioka}},\ and\ \bibinfo {author} {\bibfnamefont
  {S.-C.}\ \bibnamefont {Zhang}},\ }\bibfield  {title} {\bibinfo {title}
  {Spontaneous interlayer coherence in double-layer quantum hall systems:
  Charged vortices and kosterlitz-thouless phase transitions},\ }\href
  {https://doi.org/10.1103/PhysRevB.51.5138} {\bibfield  {journal} {\bibinfo
  {journal} {Phys. Rev. B}\ }\textbf {\bibinfo {volume} {51}},\ \bibinfo
  {pages} {5138} (\bibinfo {year} {1995})}\BibitemShut {NoStop}%
\bibitem [{\citenamefont {{Wang}}\ \emph {et~al.}(2024)\citenamefont {{Wang}},
  \citenamefont {{Fan}}, \citenamefont {{Dai}},\ and\ \citenamefont
  {{Zaletel}}}]{excitonJJ}%
  \BibitemOpen
  \bibfield  {author} {\bibinfo {author} {\bibfnamefont {T.}~\bibnamefont
  {{Wang}}}, \bibinfo {author} {\bibfnamefont {R.}~\bibnamefont {{Fan}}},
  \bibinfo {author} {\bibfnamefont {Z.}~\bibnamefont {{Dai}}},\ and\ \bibinfo
  {author} {\bibfnamefont {M.~P.}\ \bibnamefont {{Zaletel}}},\ }\bibfield
  {title} {\bibinfo {title} {{Designing exciton-condensate Josephson junction
  in quantum Hall heterostructures}},\ }\href
  {https://doi.org/10.48550/arXiv.2409.19059} {\bibfield  {journal} {\bibinfo
  {journal} {arXiv e-prints}\ ,\ \bibinfo {eid} {arXiv:2409.19059}} (\bibinfo
  {year} {2024})},\ \Eprint {https://arxiv.org/abs/2409.19059}
  {arXiv:2409.19059 [cond-mat.mes-hall]} \BibitemShut {NoStop}%
\bibitem [{\citenamefont {Hu}\ \emph {et~al.}(2023{\natexlab{b}})\citenamefont
  {Hu}, \citenamefont {He},\ and\ \citenamefont {Zhu}}]{Hu23}%
  \BibitemOpen
  \bibfield  {author} {\bibinfo {author} {\bibfnamefont {L.}~\bibnamefont
  {Hu}}, \bibinfo {author} {\bibfnamefont {Y.-C.}\ \bibnamefont {He}},\ and\
  \bibinfo {author} {\bibfnamefont {W.}~\bibnamefont {Zhu}},\ }\bibfield
  {title} {\bibinfo {title} {{Operator Product Expansion Coefficients of the 3D
  Ising Criticality via Quantum Fuzzy Spheres}},\ }\href
  {https://doi.org/10.1103/PhysRevLett.131.031601} {\bibfield  {journal}
  {\bibinfo  {journal} {Phys. Rev. Lett.}\ }\textbf {\bibinfo {volume} {131}},\
  \bibinfo {pages} {031601} (\bibinfo {year} {2023}{\natexlab{b}})}\BibitemShut
  {NoStop}%
\end{thebibliography}%

\newpage 
\cleardoublepage 

\setcounter{equation}{0}
\setcounter{figure}{0}
\setcounter{table}{0}
\setcounter{page}{1}
\setcounter{section}{0}
\makeatletter
\renewcommand{\theequation}{S\arabic{equation}}
\renewcommand{\thefigure}{S\arabic{figure}}
\renewcommand{\thesection}{S\Roman{section}}
\renewcommand{\thepage}{\arabic{page}}
\renewcommand{\thetable}{S\arabic{table}}

\onecolumngrid

\begin{center}
\textbf{\large Supplemental Online Material for ``Conformal scalar field theory from Ising tricriticality on the fuzzy sphere" }\\[5pt]
\vspace{0.1cm}
\begin{quote}
{\small In this Supplementary Material, we provide a review of the conformally coupled scalar field theory and the Hartree-Fock approximation. We also give details of the parameter optimization and discuss the pitfalls associated with false-positive variational minima. Finally, we provide additional numerical evidence for the tricritical point and the nature of phase transitions. 
}  \\[20pt]
\end{quote}
\end{center}

\section{Review of the conformally coupled scalar field theory}
\label{app:free scalar}

Let $\phi$ denote a real scalar field. The Euclidean action of the conformally invariant scalar field theory in $d>2$ dimensions, also called conformally coupled scalar, is
\begin{equation}
    S[\phi] = \frac{1}{2} \int d^d x \sqrt{\det g} \Big( (\partial \phi)^2 + \frac{d-2}{4(d-1)} R(g) \phi^2 \Big)\,,
\end{equation}
where $R(g)$ is the scalar curvature and is added as an improvement term.
In this section, we collect basic properties of the three-dimensional conformal scalar that are relevant to our discussions in the main text. We first consider the Euclidean geometry and review its operator content for $\ell \leq 4$ and $\Delta \leq 7$. We then consider the spherical geometry and review its canonical quantization.

\subsection{Conformal scalar on the plane: operator content}

In Euclidean space $\bbR^3$, the action takes a simple form 
$$
S[\phi] = \frac{1}{2} \int d^3 x (\partial\phi)^2\,.
$$ 
It follows that the scalar $\phi$ has the scaling dimension $\Delta_\phi = 1/2$, which saturates the unitarity bound in three dimensions.
Modulo the equations of motion $\partial^2 \phi = 0$, a complete basis of local operators consist of the ``unordered words" of $\phi$ and their spatial derivatives, $\phi^n$, $\partial_{\mu}\phi^n$, $\phi \partial_{\mu_1} \phi \partial_{\mu_2} \phi$, etc.,  each of which is normal ordered in order to have finite correlation functions.
Any nonzero operator that involves $n$ $\phi's$, $k$ pairs of contracted derivatives and $\ell$ non-contracted ones has a scaling dimension $\Delta = n + k + \ell$ and spin $\ell$.

Among all the local operators, some are primaries (e.g., $\phi^n$) while others are descendants (e.g., $\partial_\mu \phi)$. 
We can determine all primaries recursively as follows:
\begin{enumerate}
\item Consider all operators that are made of $n$ $\phi$'s and suppose that we know all primary operators of spin less than $\ell$ (i.e., operators involving less than $\ell$ derivatives).

\item To determine spin-$\ell$ primaries, we first list all operators with $\ell$ derivatives, schematically
\begin{equation}
    X_{n,\ell} = \partial^{\ell_1} \phi \partial^{\ell_2} \phi \ldots \partial^{\ell_i} \phi\,,\quad \ell_1 + \ldots + \ell_i = \ell,
\end{equation}
where all derivatives are non-contracted since having a pair of contracted derivatives implies it is at least a level-2 descendant of another primary. 
Furthermore, we symmetrize all the indices and subtract all traces so that the operator falls into an irreducible representation of the spatial $SO(3)$ rotation group.
For example, there are two such operators for $n=2$ and $\ell=2$
\begin{equation}
    \partial_\mu \phi \partial_\nu \phi - \frac{g_{\mu\nu}}{d} (\partial \phi)^2\,,\quad \phi \partial_\mu \partial_\nu \phi\,.
\end{equation}
However, not both of them are primary.

\item To search for new primaries from these operators, we find their linear combinations that are orthogonal to the (descendants of) known primaries by computing the overlap of the corresponding states. Specifically, we constrain the operators to the $xy$ plane and work with the complex coordinate $w = x + i y$. The fact that $g_{ww} = g_{\bar{w} \bar{w}} = 0$ implies that the purely holomorphic and anti-holomorphic components $O_{w_1,w_2,\ldots,w_\ell}$, $O_{\bar{w}_1,\bar{w}_2,\ldots,\bar{w}_\ell}$  only involve non-contracted derivatives.
Specifically, let $O_{\ell'}$ denote a known primary and $X_\ell$ a candidate of spin-$\ell$ primary. 
We define the following modified inner product
\begin{equation}
    (O_{\ell'}, X_\ell) := \lim_{|w| \rightarrow \infty} w^{\ell - \ell'} |w|^{2\Delta_{O_\ell'}} \braket{O_{\ell'}^\dag(w,\bar{w}) X_\ell(0)}\,,
\end{equation}
where $w^{\ell - \ell'}$ means that we consider the descendant of $O_{\ell'}$ that has the same energy and spin as $X_\ell$ so that the overlap does not trivially vanish.
\end{enumerate}
One can program the above recursive procedure and solve for all primaries numerically. Let us list a few results for primaries with $\ell \leq 4$:
\begin{itemize}
\item One $\phi$: no spinful primary
\item Two $\phi$'s: we get the stress-energy tensor at $\ell=2$ and another primary at $\ell=4$
\begin{equation}
    \partial_\mu\phi \partial_\nu \phi - \frac{1}{3} \phi \partial_\mu\partial_\nu \phi\,,\quad
    \partial_\mu\partial_\nu\phi \partial_\rho\partial_\sigma \phi - \frac{4}{5} \partial_\mu \phi \partial_\nu\partial_\rho \partial_\sigma \phi + \frac{1}{35} \phi \partial_\mu \partial_\nu\partial_\rho \partial_\sigma \phi
\end{equation}
\item Three $\phi$'s: we get one primary at $\ell=2$ and another at $\ell=3$
\begin{equation}
    \phi \partial_\mu\phi \partial_\nu \phi - \frac{1}{3} \phi^2 \partial_\mu\partial_\nu \phi\,,\quad
    \partial_\mu\phi\partial_\nu\phi \partial_\rho\phi - \frac{1}{2} \phi\partial_\mu\partial_\nu\phi \partial_\rho\phi+ \frac{1}{30} \phi^2\partial_\mu\partial_\nu\phi 
\end{equation}
\end{itemize}
We see that the spectrum at the tricritical point correctly captures all these states.

\subsection{Conformal scalar on the sphere: canonical quantization}

Consider the case where the spatial manifold is a sphere of radius $R$. We use $\tau\, (t)$ for the Euclidean (Lorentzian) time, $(\theta,\varphi)$ for the spherical coordinate so that the metric is $ds^2 = d\tau^2 + R^2 (d\theta^2 + \sin^2 \theta d\varphi^2)$. The Euclidean action reads
\begin{equation}
    S[\phi] = \frac{1}{2}\int d^3 x \sqrt{\det g} \big( (\partial \phi)^2 + \frac{1}{4R^2} \phi^2 \big)  \,.
\end{equation}
In this case, the equation of motion is modified by the curvature
\begin{equation}
    \partial_t^2 \phi + \frac{1}{R^2} \big( -\nabla_\Omega^2 + \frac{1}{4} \big) \phi = 0
\end{equation}
where $\nabla_\Omega$ stands for the spatial derivative on a unit sphere. The eigensolutions are given by the spherical Harmonics $Y_{\ell,m}$ with the frequency $\omega_\ell = \big( \ell + \frac{1}{2} \big)/R$.
To canonically quantize the theory, we define the canonical momentum $\Pi \equiv \partial_t \phi$ with the following commutation relation
\begin{equation}
    [\Pi(\theta,\varphi), \phi(\theta',\varphi')] = -\frac{i}{R^2 \sin\theta} \delta(\theta - \theta') \delta(\phi - \phi')\,.
\end{equation}
The Hamiltonian reads
\begin{equation}
    H = \frac{R^2}{2} \int d\phi d\theta \sin\theta \big( \pi^2 + \frac{1}{R^2} (\nabla_\Omega \phi)^2 + \frac{1}{4R^2} \phi^2 \big)
\end{equation}
To solve the energy spectrum of the theory, let us perform the mode expansion in terms of the eigensolutions of the equation of motion
\begin{equation}
    \phi = \frac{1}{R} \sum_{\ell,m} \frac{1}{\sqrt{2\omega_\ell}} \big( e^{-i \omega_{\ell} t} Y_{\ell,m} \phi_{\ell,m} + e^{i \omega_{\ell} t} Y_{\ell,m}^* \phi_{\ell,m}^\dag \big) \,,\quad \omega_{\ell} = \frac{1}{R} \big( \ell + \frac{1}{2} \big),
\end{equation}
where the $1/R$ prefactor is added as a convenient normalization factor. That $\phi \propto R^{-1/2}$ is another manifestation of its scaling dimension $\Delta_\phi = 1/2$.
One can then verify that the normal modes $\phi_{\ell,m}$ satisfy the following standard Heisenberg algebra 
\begin{equation}
    [\phi_{\ell,m}, \phi_{\ell',m'}^\dag] = \delta_{\ell,\ell'} \delta_{m,m'}\,.
\end{equation}
Accordingly, the quantum Hamiltonian reads
\begin{equation}
    H = \sum_{\ell,m} \omega_{\ell} \phi_{\ell,m}^\dag \phi_{\ell,m}\,,
\end{equation}
where the Casimir energy, the constant piece, turns out to vanish under the zeta function regularization.

\section{Details of the Hartree-Fock approximation}
\label{app:Hartree Fock}

In this section, we briefly review the Hartree-Fock approximation for the quantum Hall ferromagnet. For the purpose of this work, we only consider the ground state energy of a uniform state within this approximation. Our presentation closely follows Refs.~\cite{moon199513SpontaneousInterlayer,excitonJJ} and tailors it for the fuzzy sphere.

It is easier to state the Hartree-Fock approximation in a more general setup. Consider an electron gas confined in the lowest Landau level (LLL) with multiple flavors labeled by $a$.
The full Hamiltonian contains a single-particle part $H_0$ and the two-particle interaction $H_{\text{int}}$:
\begin{equation}
\begin{aligned}
	H =& H_0 + H_{\text{int}}, \\
	H_0 =& \int h_{ab} c_a^\dag(\r) c_b(\r) d^2\r, \\
	H_{\text{int}} =& \int V_{abcd} (\r_1 - \r_2) c_a^\dag(\r_1) c_b^\dag(\r_2) c_c(\r_2) c_d(\r_1) d^2\r_1 d^2\r_2 \,.
\end{aligned}
\end{equation}
where $c_a(\r)$ is the LLL-projected fermion annihilation operator and $V_{abcd}(\bm{r})$ is a generic two-body interaction.
We can recast the Hamiltonian in orbital space as
\begin{equation}
\begin{aligned}
	H_0 =& \sum_{m}h_{ab} c_{m,a}^\dag c_{m,b}, \\
	H_{\text{int}} =& \sum_{\substack{m_1,m_2,\\m_3,m_4}} V^{m_1,m_2,m_3,m_4}_{abcd} \, \delta_{m_1+m_2,m_3+m_4}\, c_{m_1,a}^\dag c_{m_2,b}^\dag c_{m_3,c} c_{m_4,d},  \\
	V^{m_1,m_2,m_3,m_4}_{abcd} = & V_{abcd} \sum_J V_J (4q-2J+1) 
	\begin{pmatrix} q & q & 2q-J \\ m_1 & m_2 & -m_1-m_2 \end{pmatrix}
	\begin{pmatrix} q & q & 2q-J \\ m_4 & m_3 & -m_3-m_4 \end{pmatrix},
\end{aligned}
\end{equation}
where $N_{\phi}=2q$ is the number of flux quanta, the LLL orbitals are labeled by indices $m_i = -q,\ldots,q$, the interaction matrix elements $V_{abcd}$ depend only on the flavor indices and $V_{J=0,1,\ldots,2q}$ are the Haldane pseudopotentials. 

Within the Hartree-Fock approximation, the ground state wave function is a Slater determinant and we can capture all many-body properties by the fermion two-point correlation functions.
Assuming a spatially uniform ground state, the fermion two-point function is
\begin{equation}
	\langle c^\dag_{m,a}c_{m',b}\rangle = \delta_{m,m'} P_{ab}\,, 
\end{equation}
where $P^2 = P$ is a Hermitian spectral projector. 
The ground state energy per orbital reads:
\begin{equation}
\begin{gathered}
	\frac{E_{\text{HF}}}{N_\phi + 1} = h_{ab} P_{ab} + V^{\text{H}}_{abcd} P_{ad} P_{bc} - V_{abcd}^{\text{F}} P_{ac} P_{bd}\,, \\
	V^{\text{H}}_{abcd} = V_{abcd} \sum_J V_J \,,\quad 
	V^{\text{F}}_{abcd} = V_{abcd} \sum_J (-1)^J V_J\,.
\end{gathered}
\end{equation}
We can then combine the even and odd pseudopotentials to get:
\begin{equation}
	\frac{E_{\text{HF}}}{N_\phi + 1} = h_{ab} P_{ab} + V_{abcd} \sum_{J} V_{2J} \big( P_{ad} P_{bc} - P_{ac} P_{bd} \big) + V_{abcd} \sum_{J} V_{2J+1} \big( P_{ad} P_{bc} + P_{ac} P_{bd} \big)\,.
\end{equation}
We now apply the general result to the bilayer quantum Hall system in the main text. There are two flavors $a = \uparrow, \downarrow$. The intralayer interaction $U$ corresponds to $a = b = c = d = \uparrow,\downarrow$, and the interlayer interaction $V$ corresponds to $a = d = \uparrow, b = c = \downarrow$. At unit filling, we parameterize the spectral projector $P$ by a unit vector $\bm{m}$ as $P = (1 + \bm{m} \cdot \bm{\sigma})/2$. The Hartree-Fock energy per particle of the Hamiltonian in Eq.~\eqref{eq:TCI fuzzy} of the main text is
\begin{equation}
	\frac{E_{\text{HF}}}{N} = - h m_x + \sum_J (\lambda U_{2J+1} - V_{2J+1}) m_z^2 + \const\,,
\end{equation}
where $U_J$ and $V_J$ are the coefficients for the pseudopotential expansion of the intralayer and interlayer interaction. 
Without loss of generality, we restrict to the $m_y = 0$ plane and take $m_z$ as the order parameter for the ferromagnetic phase.
Near the phase transition ($|m_z| \ll 1$), we can write the energy per flux as
\begin{equation}
\begin{aligned}
	\frac{E_{\text{HF}}}{N_\phi} = \big( \lambda U_{\text{intra}} - V_{\text{inter}} + \frac{h}{2} \big) m_z^2 + h \big( \frac{1}{8} m_z^4 + \frac{1}{16} m_z^6 + \ldots \big)\,,
\end{aligned}
\end{equation}
where $U_{\text{intra}} = \sum_J U_{2J+1}$, $V_{\text{inter}} = \sum_J V_{2J+1}$, and both are chosen to be positive.
In particular, the coefficients of higher-order terms are all proportional to the transverse field $h$, which allows us to determine the phase boundary simply via the vanishing of the quadratic term as
$2 \lambda U_{\text{intra}} + h = 2 V_{\text{inter}}$.
For any $h>0$, the higher-order terms are all positive. The transition is second-order and it falls into the Gaussian universality class.
At $h = 0$, all higher-order terms vanish and the transition becomes first-order.

\section{Variational optimization and ``fake'' solutions}

In this section, we provide details of the gradient descent optimization procedure. We first explain the construction of our cost function for targeting the desired conformal point and then discuss the ``fake" solutions that can arise with  naively defined cost functions, highlighting the delicacy of this approach. 

\subsection{Details of the cost function}

Our gradient descent procedure was performed for a fixed size of $N=10$ electrons over all states with spin-$\ell\leq 2$ and theoretically expected scaling dimensions $\Delta<3$, as well as the second spin-2, $\mathbb{Z}_2$-even state. For the free scalar CFT, we expect two spin-2 states with $\Delta=3$ -- one corresponding to $T_{\mu\nu}$ and the other to $\partial_\mu\partial_\nu\phi^2$. As highlighted in \figref{fig:nx exp value} below, it is difficult to distinguish these states through simple means. Therefore, we normalize the first spin-2, $\mathbb{Z}_2$-even state -- which we will call $O_{\mu\nu}$ -- to have $\Delta=3$ and include the second in our optimization procedure. To define the cost function, let us denote by $\boldsymbol{\Delta} = (\Delta_\phi,\Delta_{\phi^2},...,\Delta_{O_{\mu\nu}})$ and $\mathbf{E} = (E_1,E_2,...,3)$ the vectors containing the expected scaling dimensions and rescaled energies of the states considered, respectively.
Note that, as we have normalized $\Delta_{O_{\mu\nu}}=3$, the last element of each vector always equal each other. Nevertheless, we keep them here in order to avoid the ambiguity of overall rescaling, since without fixed normalization we could have solutions where $\mathbf{E} = \kappa\boldsymbol{\Delta}$ with any $\kappa \neq 1$. Using the expression for the length of the component of a vector $\mathbf{v}$ that is orthogonal to $\mathbf{u}$:
\begin{equation}
    |\textrm{Orth}_{\mathbf{u}}(\mathbf{v})|^2 = |\mathbf{v}|^2 - \frac{(\mathbf{u} \cdot \mathbf{v})^2}{|\mathbf{u}|^2},
\end{equation}
the spectral cost function reads
\begin{equation}
f_s(\{U,V\},h) = |\textrm{Orth}_{\mathbf{E}}(\boldsymbol{\Delta})|^2,
\label{eq:naive cost function}
\end{equation}
where $\mathbf{E}$ implicitly depends on the pseudopotentials, $U,V$, and the transverse field $h$.

In practice, we find that using the cost function (\ref{eq:naive cost function}) can easily result in local minima that correspond to ``fake" tricritical Ising spectra -- a numerical quirk which does not occur when the same analysis is done for the 3D Ising CFT. At these local minima, the spectra agree well with that of the free-scalar CFT at low energies, but they are in fact gapped, with exponentially decaying correlations. We will return to the ``fake" solutions later as they are interesting in their own right. To systematically eliminate the ``fake" minima of our spectral cost function, we supplement the cost function by the two-point, equal-time correlator, $G_{\phi\phi}(r=1,\theta=\pi)$, evaluated between the poles of the sphere. 
For the free scalar CFT, we have~\cite{Han:2023yyb,HuOPE2023}
\begin{equation}
\label{eqn: free correlator}
    G^\text{free}_{\phi\phi}(r=1,\theta) = \frac{1}{2\sin(\theta/2)}, 
\end{equation}
where we assumed that  $n_z(\Omega)$ serves as a good approximation for our lowest lying $\mathbb{Z}_2$-odd operator, $\phi$, at the critical point \cite{HuOPE2023}.  
It is clear from \eqnref{eqn: free correlator} that $G^\text{free}_{\phi\phi}(r=1,\pi) = 0.5$. Hence, in our modified cost function we use the same expression (\ref{eq:naive cost function}) with modified vectors
\begin{equation}
\begin{aligned}
    \boldsymbol{\tilde{\Delta}} &=(0.5,\boldsymbol{\Delta}),\\
    \boldsymbol{\tilde{E}} &=(G_{\phi\phi}(r=1,\pi),\boldsymbol{E}).   
\end{aligned}
\end{equation}
Through a numerical optimization procedure, with $N=10$ electrons and $U_1 = V_1 = 1$ fixed, we obtain the following Hamiltonian parameters:
\begin{equation}\label{eq:optparamfull}
\begin{gathered}
	\{ U_1,U_3,U_5 \} = \{1.0, 0.06838529084307, -0.13373096252546 \}\,, \\
	\{ V_{0},V_{1},V_{2} \} = \{2.346948983144344,1.0, 0.37295992394339 \}\,,\\
    h = 0.23092194571912\,.
\end{gathered}
\end{equation}

\subsection{The Ising line}

We now illustrate the determination of the Ising transition line in the main text. A simple cost function that quantifies the deviation of the low energy spectrum from the Ising CFT spectrum is
\begin{equation}
	f(\lambda,h) = (\alpha E_{\lambda,h}^\sigma - \Delta_\sigma)^2\,, 
	\label{eq:cost function Ising line}
\end{equation}
where $E_{\lambda,h}^\sigma$ is the numerically obtained energy of the lowest state in the $\bbZ_2$-odd sector, $\Delta_\sigma\approx 0.5181489$ is the scaling dimension of the $\sigma$ primary in the 3D Ising CFT~\cite{Poland19}, and $\alpha$ is a rescaling factor chosen such that the first spin-$2$, $\mathbb{Z}_2$-even state lies at energy $3$. 

\begin{figure}
\centering
\includegraphics[width = 0.75\textwidth]{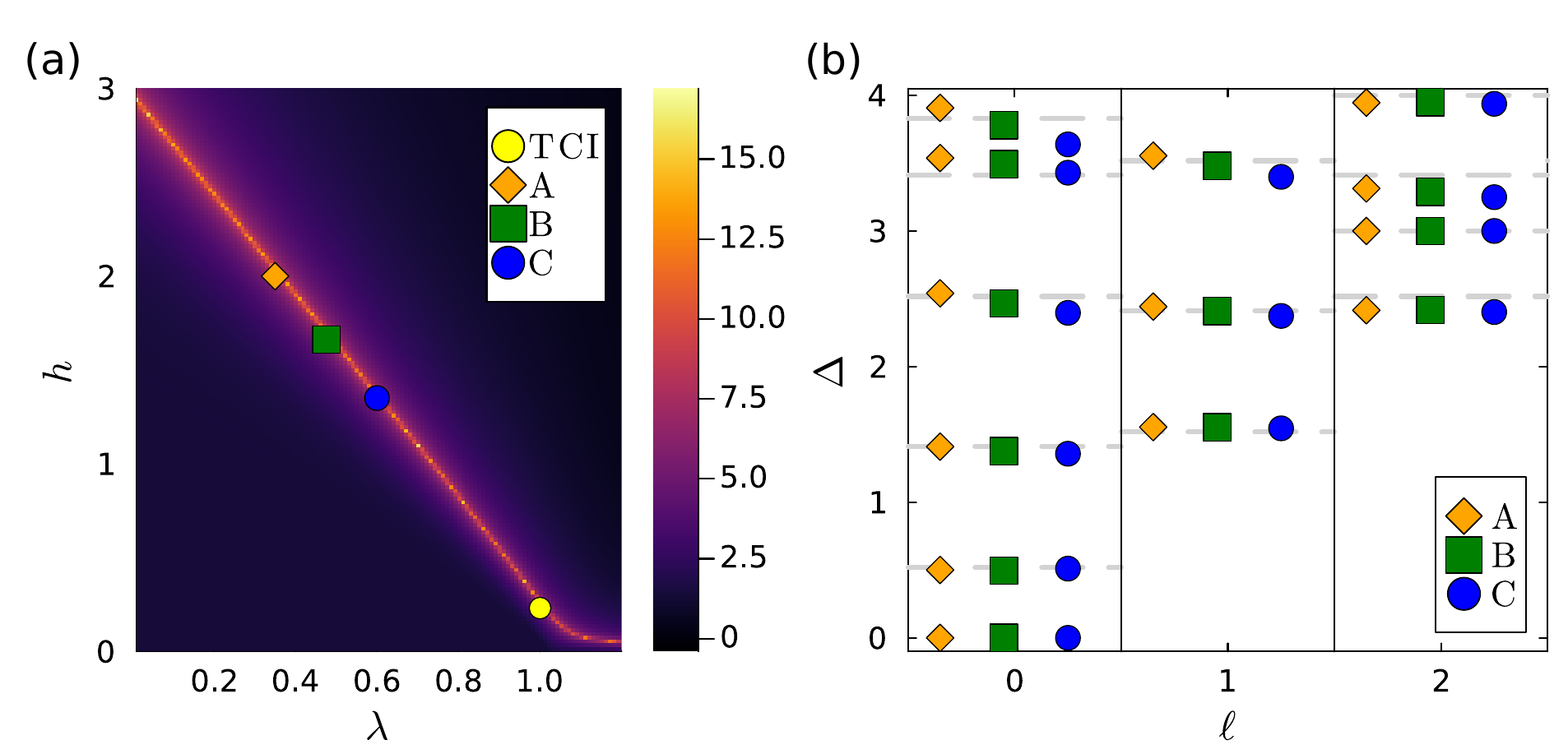}
\caption{(a) The negative logarithm of the cost function, Eq.~(\ref{eq:cost function Ising line}), across the phase diagram. The minimum of the cost function (bright line) defines the 3D Ising transition line, which was plotted in Fig.~\ref{fig:phase diagram}(a) of the main text. The 3D Ising line terminates at the tricritical point (yellow dot). Below the yellow dot, the minimum value grows with system size and we attribute this feature to a finite-size effect. 
(b) The low-energy spectra at three distinct points on the Ising line in $(\lambda,h)$-space: $A = (0.35,2.0)$; $B=(0.476,1.66)$ and $C=(0.60,1.35)$. The dashed lines are the expectations based on the 3D Ising CFT. All data is for $N=12$ electrons.}
\label{fig:Ising Line}
\end{figure}

We evaluate the cost function $f(\lambda,h)$ across the phase diagram for $N=12$ electrons in \figref{fig:Ising Line}(a). 
A clear line of local minima emerges. We identify the portion of this line above the tricritical point (yellow dot) as the Ising transition line -- see \figref{fig:Ising Line}(b) for the spectra at three representative points along this line. Below the tricritical point, as can be seen in \figref{fig:Ising Line}(a), the cost function still displays a minimum. Nevertheless, upon further inspection of this region across system sizes $N=6-12$, we verified that the cost function grows with system size below the tricritical point. Therefore, we attribute the minima below the yellow dot in \figref{fig:Ising Line}(a) to finite-size effects, rather than a continuation of the critical line.

\begin{figure}
\centering
\includegraphics[width = 0.75\textwidth]{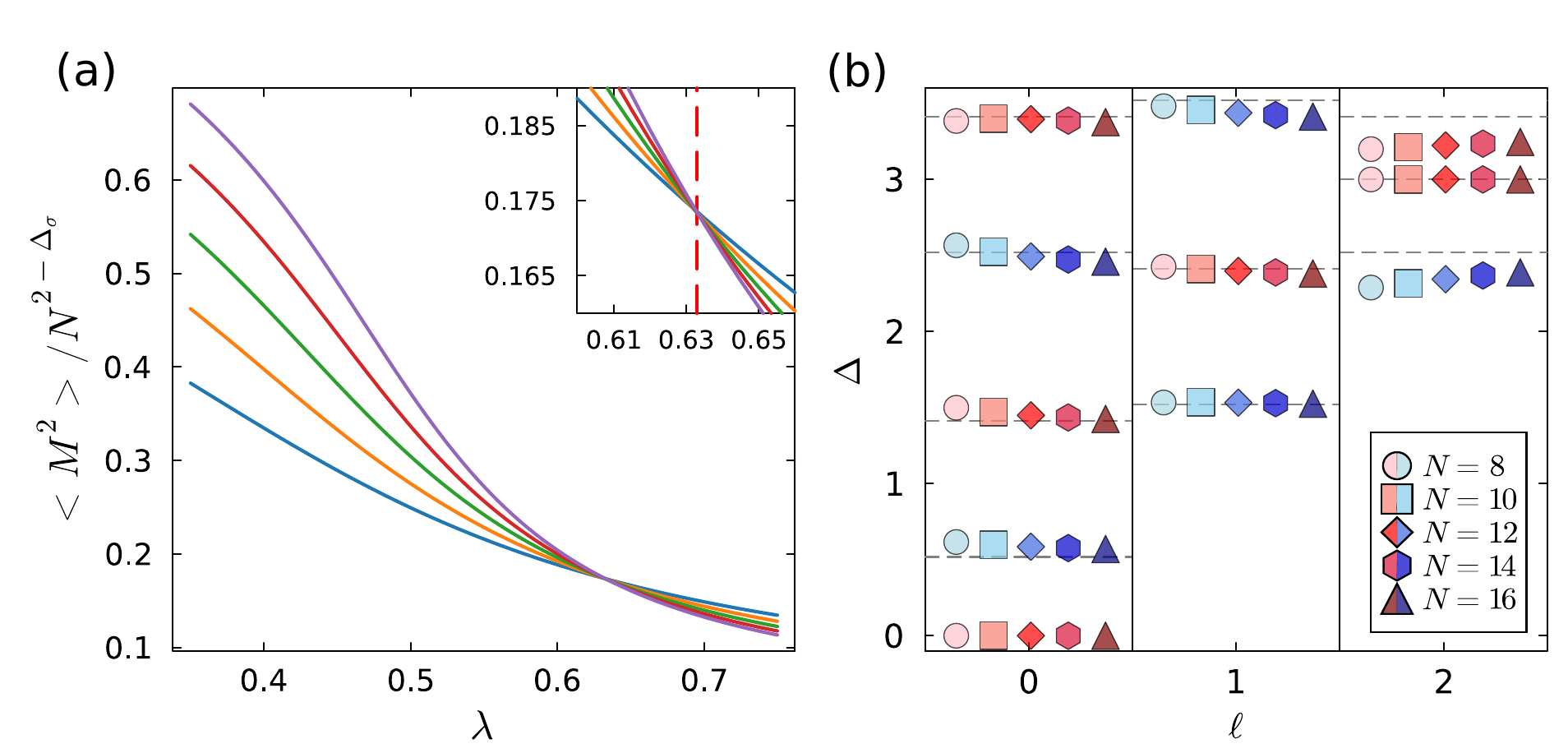}
\caption{ (a) The finite size scaling of the magnetization along a fixed $h=1.35$. (b) The low energy spectrum at the crossing point $\lambda_c \approx 0.633$, with the predicted scaling dimensions for the Ising CFT shown by dashed lines. Data is for several system sizes shown in the legend, while red and blue symbols indicate $\mathbb{Z}_2$-even and odd states, respectively.   }
\label{fig:Ising l 0.63}
\end{figure}

Furthermore, we find that the Ising transition line drifts towards larger $\lambda$ and $h$ as the system size increases, consistent with the finite-size scaling analysis presented in the main text. For instance, at $h=2.0$ (corresponding to the point $A$ above), the crossing point is found to occur at $\lambda = 0.4$ -- slightly larger than $0.35$ shown in \figref{fig:Ising Line}(a). In \figref{fig:Ising l 0.63} we perform the same analysis for $h=1.35$ (in the neighborhood of point $C$) and observe a similar drift. 

\begin{figure}
\centering
\includegraphics[width = 0.6\textwidth]{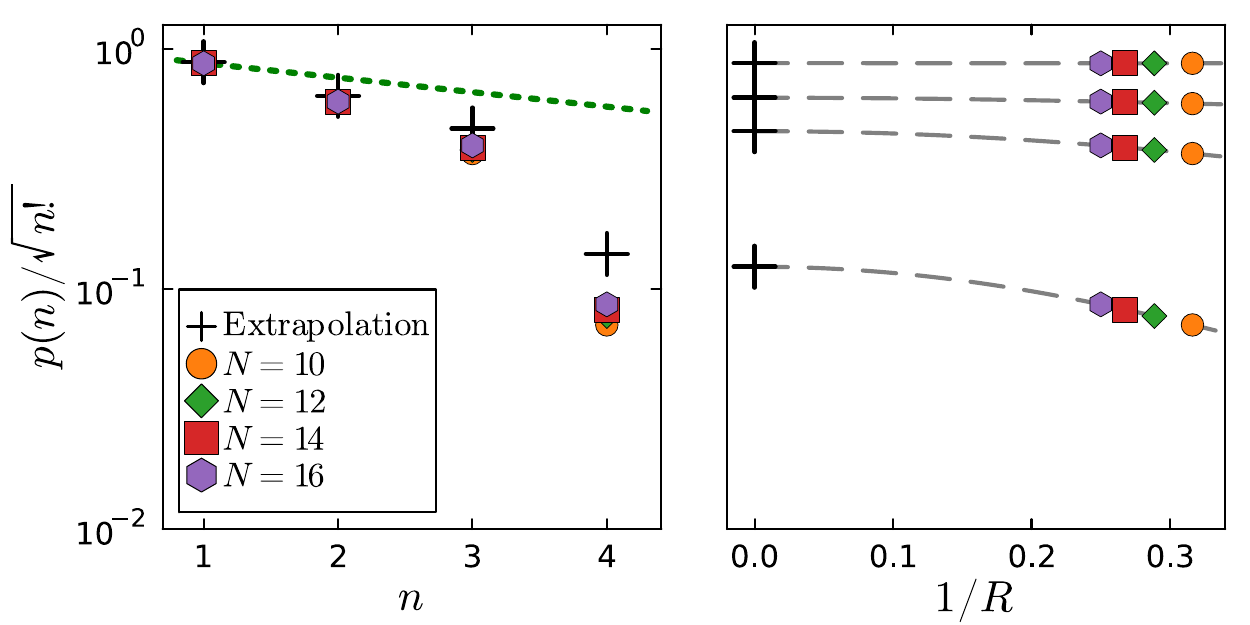}
\caption{$p(n)/\sqrt{n!}$ for different system sizes and its finite-size extrapolation at a point $(\lambda,h)=(0.4,2.0)$ along the 3D Ising line. This analysis should be contrasted against Fig.~\ref{fig:Tower} of the main text for the tricritical point. Left: The filled symbols are the numerical results and crosses are the extrapolated values. The dashed line indicates a departure from the linear dependence with $n$ on a log scale. Right: finite-size extrapolation is performed with two terms, $1/R^2$ and $1/R^{4.77}$, according to \eqnref{eqn: Ising Nz}. 
}
\label{fig: Ising tower}
\end{figure}

Finally, we compute the equivalent of \figref{fig:Tower} in the main text, which will allow us to draw a direct comparison between the 3D Ising line and the tricritical point. We choose the same point $(\lambda,h)=(0.4,2.0)$ as for the 3D Ising spectra in \figref{fig:phase diagram}(b) of the main text. We now define our $\widetilde{N}_z$ for the Ising CFT as 
\begin{equation}
\begin{aligned}
\label{eqn: Ising Nz}
    \widetilde{N}_z = \frac{N_z}{R^{2 - \Delta_\sigma}} &= a_1 \sigma_{0,0} + \frac{a_2}{R^{\Delta_{\Box\sigma}-\Delta_{\sigma}}} (\Box\sigma)_{0,0}  + \frac{a_3}{R^{\Delta_{\sigma'}-\Delta_{\sigma}}} \sigma^\prime_{0,0} + \ldots \\
    &\approx a_1 \sigma_{0,0} + \frac{a_2}{R^{2}} (\Box\sigma)_{0,0}  + \frac{a_3}{R^{4.77}} \sigma^\prime_{0,0}  + \ldots  
\end{aligned}
\end{equation}
where $\hat{O}_{0,0}$ is the $(l,m)=(0,0)$ spherical mode of the operator $\hat{O}$, discussed in Ref.\cite{Hu23} for Ising CFT operators. Analogously to \eqnref{eqn:free boson}, we define 
\begin{equation}
p(n) = \bra{n}\widetilde{N}_z^n\ket{0},
\end{equation}
where $\ket{n}$ is the $n^{th}$ excited state with $\ell=0$. \figref{fig: Ising tower} shows $p(n)$ plotted against $n$, where we have now extrapolated using \eqnref{eqn: Ising Nz}. Note that, we do not consider the 5-th state for the Ising CFT, $\ket{\varepsilon^\prime}$, as the Ising $\mathbb{Z}_2$ symmetry requires $p(5) = 0$. The finite size and extrapolated values visibly deviate from the relationship given by \eqnref{eqn:free boson}. 

\subsection{``Fake" solutions}

Running the gradient descent algorithm with the most straightforward choice of cost function,  Eq.~\eqref{eq:naive cost function}, often leads to ``fake" solutions, where the system remains gapped despite exhibiting an energy spectrum that superficially resembles that of the free scalar CFT. Here, we analyze in more detail these curious false positives.

\begin{figure}
\centering
\includegraphics[width = 0.74\textwidth]{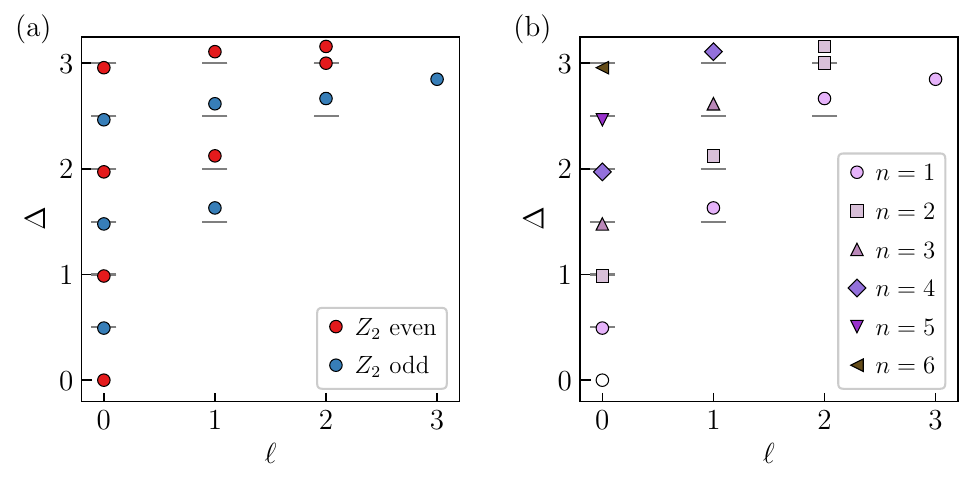}
\caption{Rescaled energy spectrum at the fake parameter. Dots are numerical values and lines are CFT expectations. (a) Energies labeled by the $\bbZ_2$ symmetries of the corresponding eigenstates (b) Energies labeled by the $U(1)$ quantum number of the corresponding eigenstates. We choose $N = 10$, $h = 1$, $V_0 = 0.84$, $V_1 = 0$, $V_2 = 0.66$.}
\label{fig:fake spectrum}
\end{figure}

For example, at $N=10$ electrons, one such “fake” solution is to set all intralayer interactions to zero ($U_{J} = 0$) while the interlayer interactions are $V_0 = 0.85, V_1 = 0, V_2 = 0.66$, with transverse field $h = 1$. The spectrum for these parameters is shown in \figref{fig:fake spectrum}(a). 
After rescaling, most of the eigenenergies align closely with integer and half-integer values—just as one would expect for a free scalar CFT.
At first glance, the operator content appears remarkably consistent with that of the free scalar theory.
Notably, we observe a unique $\bbZ_2$-odd state at $\ell=0$, $\Delta = 5/2$, two $\bbZ_2$-even state at $\ell = 2$, $\Delta = 3$, both matching the expected low-lying spectrum of the free scalar CFT.
However, a clear discrepancy emerges in the $\bbZ_2$ even sector at $\ell=0$ and $\Delta = 3$ where we find only a single state, whereas the free scalar CFT predicts two.

To definitively rule out this "fake" solution, we note a special property of the corresponding parameter choice: the interaction part possesses one accidental $SO(3)$ symmetry.
Specifically, let us parameterize the interalayer interaction with the Haldane pseudo-potentials
\begin{equation}
\begin{gathered}
    \int V(\boldsymbol{\Omega}_{1}-\boldsymbol{\Omega}_{2}) n_\uparrow (\boldsymbol{\Omega}_1) n_\downarrow(\boldsymbol{\Omega}_2) d^2 \boldsymbol{\Omega}_1 d^2 \boldsymbol{\Omega}_2 = \sum_{\substack{m_1,m_2,\\m_3,m_4}} V_{m_1,m_2,m_3,m_4} c_{\uparrow,m_1}^\dag c_{\downarrow,m_2}^\dag c_{\downarrow,m_3} c_{\uparrow,m_4}\,, \\ 
    V_{m_1,m_2,m_3,m_4} = \sum_J V_J (4s-2J+1) 
    \begin{pmatrix} q & q & 2q-J \\ m_1 & m_2 & -m_1-m_2 \end{pmatrix}
    \begin{pmatrix} q & q & 2q-J \\ m_4 & m_3 & -m_3-m_4 \end{pmatrix}\,.
\end{gathered}
\end{equation}
Recall the symmetry property of the Wigner $3j$ symbol
\begin{equation}
    \begin{pmatrix} j_1 & j_2 & j_3 \\ m_1 & m_2 & m_3 \end{pmatrix} = (-1)^{j_1+j_2+j_3} \begin{pmatrix} j_2 & j_1 & j_3 \\ m_2 & m_1 & m_3 \end{pmatrix}\,.
\end{equation}
Thus, when only $V_{2J}$'s are nonzero, the interaction matrix satisfies
\begin{equation}
    V_{m_1,m_2,m_3,m_4} = V_{m_2,m_1,m_3,m_4} = V_{m_1,m_2,m_4,m_3}\,,
\end{equation}
and we can rewrite the interaction as a product of two $SO(3)$ flavor-singlets as follows
\begin{equation}
    \frac{1}{2}\sum_{\substack{m_1,m_2,\\m_3,m_4}} V_{m_1,m_2,m_3,m_4} (c_{\uparrow,m_1}^\dag c_{\downarrow,m_2}^\dag - c_{\downarrow,m_1}^\dag c_{\uparrow,m_2}^\dag ) (c_{\downarrow,m_3} c_{\uparrow,m_4} - c_{\uparrow,m_3} c_{\downarrow,m_4})\,.
\end{equation}
Therefore, the interaction is invariant under a flavor $SO(3)$ rotation. 
Adding a transverse field along the $x$-axis, the symmetry is broken down to $U(1)$ and the ground state is fully polarized:
\begin{equation}
    \ket{\Psi_0} = \prod_m \frac{c_{m,\uparrow}^\dag + c_{m,\downarrow}^\dag}{\sqrt{2}} \ket{0},
\end{equation}
where $\ket{0}$ is the Fock vacuum, confirmed by the numerics. Thus, at this fake parameter, the real-space correlation function of the ground state wave function does not support an algebraically decaying behavior.

Given the presence of the accidental $U(1)$ symmetry, we can examine the quantum numbers carried by each eigenstate, as measured by the total pseudospin operator $N_x$.
Taking the ground state to have zero $U(1)$ quantum number as a reference, the results are shown in \figref{fig:fake spectrum}(b). 
We find that the scalar states with energy around $\Delta = n/2$ for $n = 1,2,\ldots$ carry $U(1)$ charge $n$.
Similarly, spinful states with angular momentum $\ell$ and energy around $n/2 + \ell$ also has a $U(1)$ quantum number $n$.

It is then instructive to draw a comparison between  the fake solution, the optimized tricritical Ising point and the free scalar CFT from the $U(1)$ symmetry perspective.
The free scalar CFT possesses a genuine $U(1)$ symmetry, as is manifested through its Fock space structure.
As we have shown in the previous subsection, we use the total pseudospin operator $N_x$ to confirm the predicted Fock space structure at the optimized tricritical Ising point, which implies the emergence of such a $U(1)$ symmetry.
In contrast, the fake solution has an exact $U(1)$ symmetry, which also produces surprisingly a spectrum closely resembling that of the free scalar CFT. 
A natural physical explanation is that at the fake solution, the ground state is a fully polarized state, and the excitations correspond to states with $n$ pseudospins carrying definite total angular momentum. The interactions are fine-tuned so that the resulting dispersion is nearly linear, giving rise to an $U(1)$ symmetry resolved energy spectrum that mimics the CFT structure.
Nonetheless, the system remains gapped, and its spectrum does not fully match the expected operator content of a free scalar theory. Still, the close resemblance raises an intriguing possibility: the fake solution, while not itself conformal, may lie near a free scalar CFT in theory space and could potentially be deformed or embedded into a construction that realizes the genuine fixed point.

\section{Further evidence for the tricritical point}

\begin{figure}
\centering
\includegraphics[width = \linewidth]{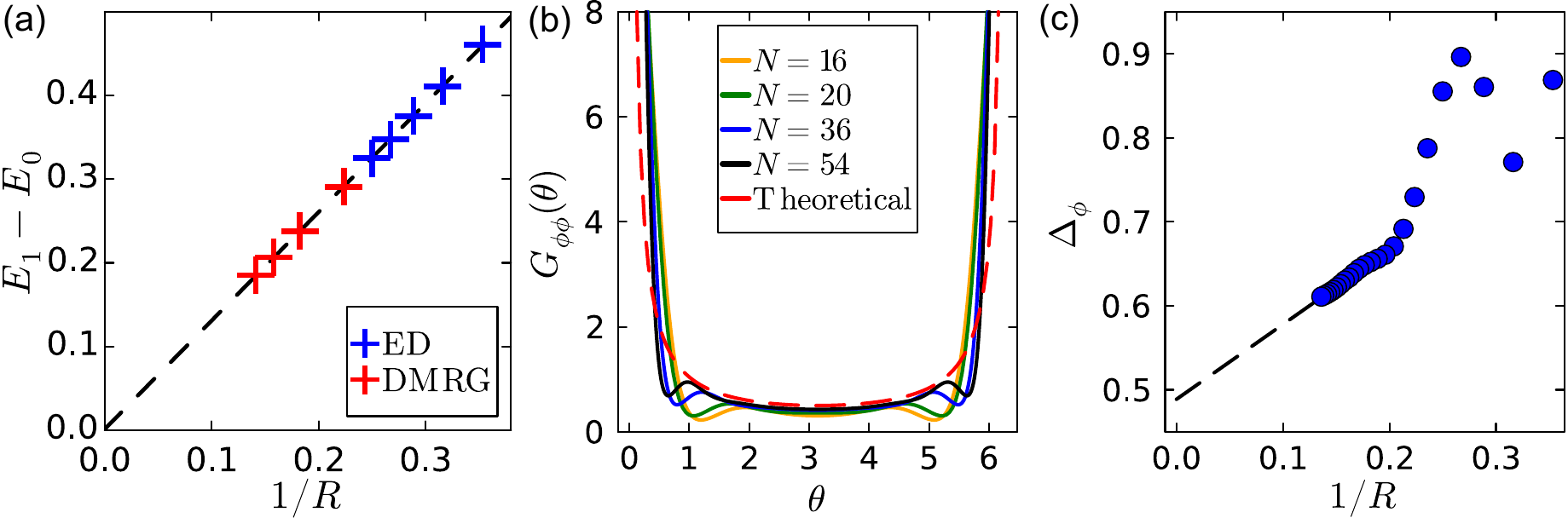}
\caption{(a) Energy gap between the ground state and first excited state extrapolates linearly in $1/R$ to a value 0.00121 close to zero. (b) The correlator $G_{\sigma\sigma}(\theta)$ for different system sizes converges to the theoretical curve, \eqnref{eqn: free correlator}, at large distances, i.e., for $\theta$ near the equator.  (c) Extrapolation of the scaling dimension calculated from the antipodal value, $G_{\phi\phi}(\theta=\pi)$.
Data for $N=8-16$ is obtained via exact diagonalization, while $18 \leq N \leq 38$ is calculated using DMRG with bond dimension 2000, and bond dimension 4000 for $40\leq N\leq 54$.
}
\label{fig:Correlator and Gap}
\end{figure}

In addition to the analysis presented in the main text, we now further validate the identification of \eqnref{eq:optparamfull} as a tricritical Ising point.
As mentioned in the main text, the energy gap between the ground and first excited state, shown \figref{fig:Correlator and Gap}(a), decreases linearly with $1/R$ up to $N=54$ electrons, with no visible deviation. 
This behavior implies that the couplings of the relevant perturbations must be small at moderate system sizes and they continue to decrease as the system grows, consistent with the flow toward the free scalar fixed point.

We further examine the behavior of the two-point correlation function $G_{\phi\phi}$ across different system sizes, as is shown in \figref{fig:Correlator and Gap}(b). Compared with 3D Ising CFT~\cite{Zhu23}, one notable difference is the presence of oscillations at small angular separations $\theta$. 
These oscillations may originate from fractional quantum Hall correlations at short distances, induced by intralayer interactions. As these features are tied to gapped degrees of freedom, they are expected to shift toward smaller angles as the system size increases—which is indeed what we observe.  
Finally, we examine the finite-size scaling of the anti-podal value of the correlator in \figref{fig:Correlator and Gap}(c), which extrapolates to a value in good agreement with the CFT prediction. The small residual deviation may be attributed to marginally irrelevant couplings. 

\begin{figure}
\centering
\includegraphics[width = 0.75\textwidth]{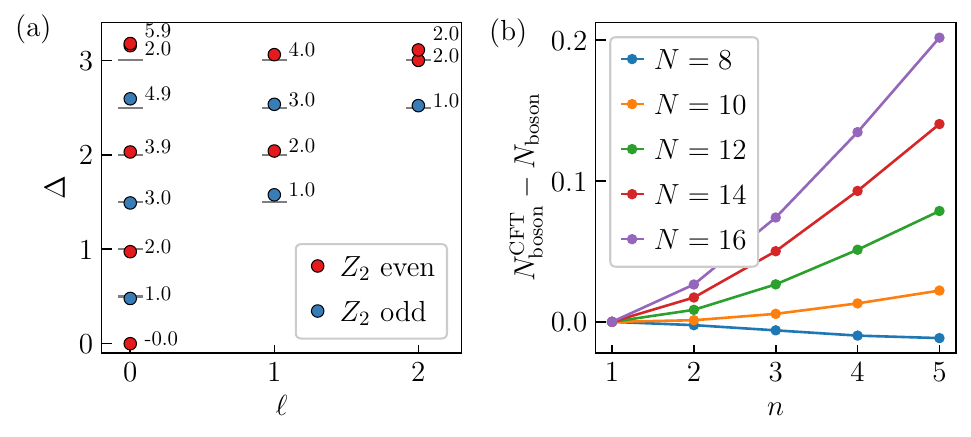}
\caption{Measured boson number of each eigenstates and its deviation from the CFT prediction. (a) We compute the boson number with $N=12$ electrons according to Eq.~(\ref{eq:Nx}) for each eigenstate. Dots are the energy and the numbers nearby are the calculated results. (b) The deviation $N_{\mathrm{boson}}^{\mathrm{CFT}} - N_{\mathrm{boson}}$ for the low-lying scalar states $\ket{\phi^n}$, $n = 1,\ldots,5$.}
\label{fig:nx exp value}
\end{figure}

As mentioned in the main text, we can further verify the emergence of the free boson algebra by analyzing the $\bbZ_2$-even total pseudospin operator $N_x$. On symmetry grounds, its general form reads
\begin{equation}
    N_x = a_0 + a_2 \sum_{\ell,m} \phi_{\ell,m}^\dag \phi_{\ell,m} + \tilde{a}_2 (\phi_0^2 + h.c.) + \tilde{a}_4 (\phi_0^4 + h.c.) + \ldots
\end{equation}
This expansion is constrained only by $\bbZ_2$ symmetry, Hermiticity and rotational invariance of the operator. We do not assume Lorentz invariance and thus treat $a_2$ and $\tilde{a}_2$ as independent numbers.
For comparison, if we do assume Lorentz invariance, the pseudo-spin density would admit an operator expansion of the form $n_x = a_0 + a_2 \phi^2 + a_4 \phi^4 + \ldots$ in terms of local CFT operators.
Given the Fock space structure, we can then use $N_x$ as a proxy for the boson number in a given energy eigenstate $\ket{E_n}$, defined as
\begin{equation}\label{eq:Nx}
    N_{\text{boson}} = \frac{\braket{E_n|N_x|E_n} - \braket{0|N_x|0}}{\braket{\phi|N_x|\phi} - \braket{0|N_x|0}},
\end{equation}
where $\ket{0}$ and $\ket{\phi}$ denote the ground state and the first excited state, respectively, and are normalized to correspond to 0 and 1 boson.
Using data for $N=12$ electrons, we show in \figref{fig:nx exp value}(a) that the extracted boson number agrees remarkably well with CFT expectations.
For instance, the $\phi$ state and its descendants all have $N_x \approx 1$. The two scalar states near $\Delta = 3$ have $N_x \approx 2$ and $N_x \approx 6$, which are in perfect agreement with identifying them as the descendant $\square \phi^2$ and the primary $\phi^6$, respectively. 

However, we observe that deviations from the ideal CFT values become more pronounced as the system size increases from $N=8$ to $N=16$ electrons, as shown in \figref{fig:nx exp value}(b). This also supports the claim in the main text that, at moderate system sizes, the system at the numerically identified tricritical point should be viewed as a free scalar theory perturbed by small but finite interactions. Indeed, one can verify that the magnitude of the observed deviations is consistent with expectations from conformal perturbation theory.
 
\section{First-order transition}

\begin{figure}
\centering
\includegraphics[width = 0.45\textwidth]{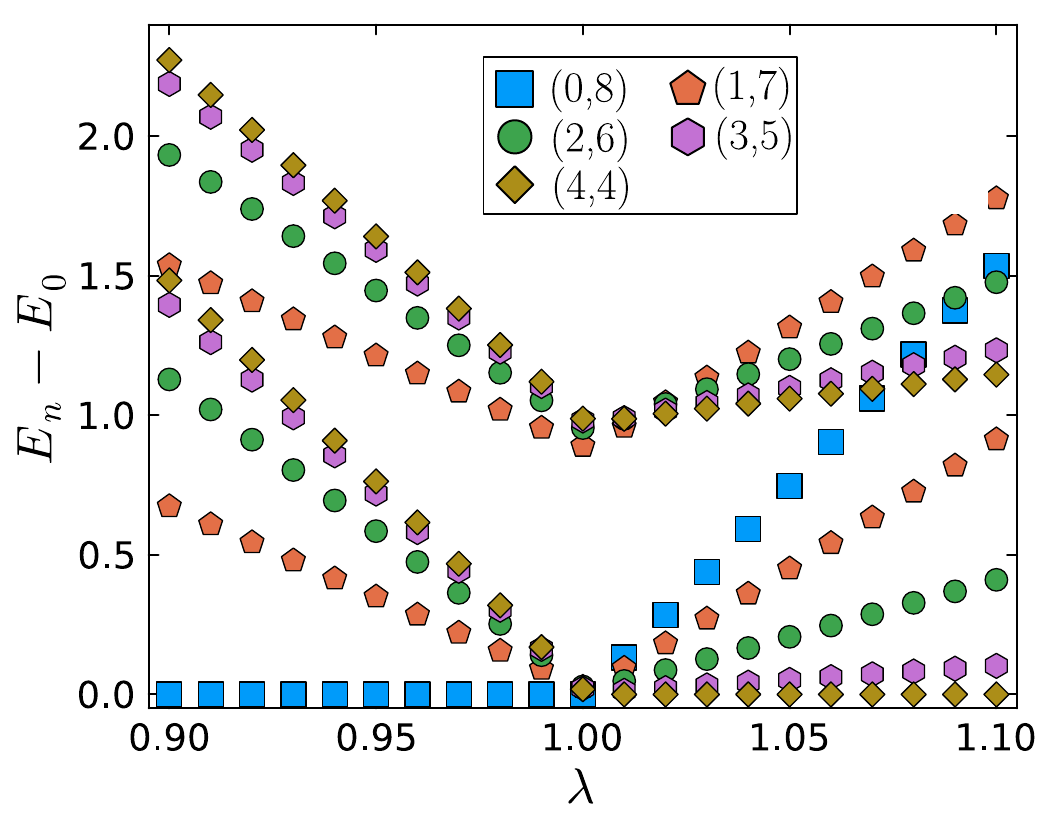}
\caption{The crossing of energy levels on the $h=0$ axis as a function of $\lambda$, for $N=8$ electrons. The levels in this case can be labeled by the numbers of electrons in each layer, $N_\uparrow$ and $N_\downarrow$, which are conserved due to a $U(1)\times U(1)$ symmetry. The spectrum in any sector $(N_\uparrow, N_\downarrow)$ is exactly the same as in  $(N_\downarrow, N_\uparrow)$ and only the two lowest energies in each sector are plotted. }
\label{fig:h_0_crossing}
\end{figure}

In this section, we examine the first-order transition in more detail. We begin with the case along the $h=0$ axis, shown in Fig.~\ref{fig:h_0_crossing}. While this case unambiguously exhibits a first-order transition, in agreement with Hartree-Fock discussed above, it contains one important subtlety: the zero-magnetisation state on the paramagnetic side is likely a charge-gapless composite Fermi liquid (CFL). This is because, at $h=0$, we have a $U(1)\times U(1)$ symmetry, with the particle number preserved in each layer. Hence, increasing $\lambda$ at $h=0$ gives rise to a level crossing between the state with one layer filled (e.g., $\nu_\uparrow=1$, $\nu_\downarrow=0$) and the state with equal density in each layer, $\nu_\uparrow=\nu_\downarrow=1/2$. The latter is expected to have gapless excitations in the charge (as well as neutral) sector, which would significantly complicate the analysis. Below we present evidence that this can be avoided with a fairly small $h>0$, such that we have a first-order transition between charge-gapped ferromagnetic and paramagnetic phases.

\begin{figure}
\centering
\includegraphics[width = 0.35\textwidth]{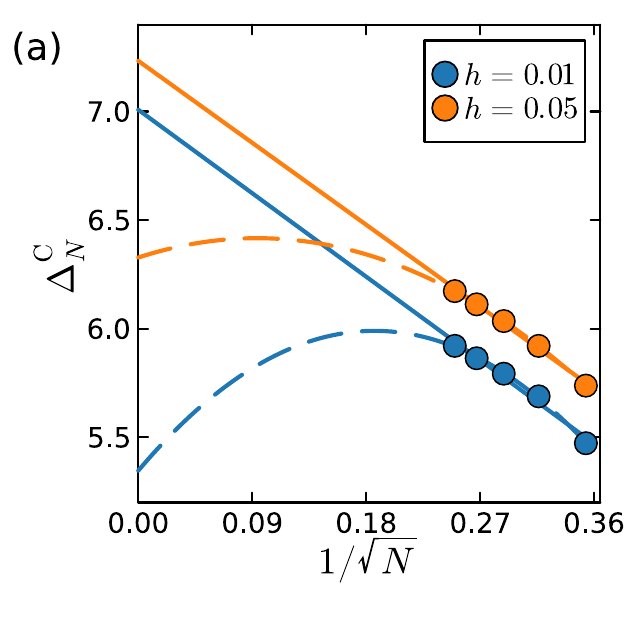}
\includegraphics[width = 0.35\textwidth]{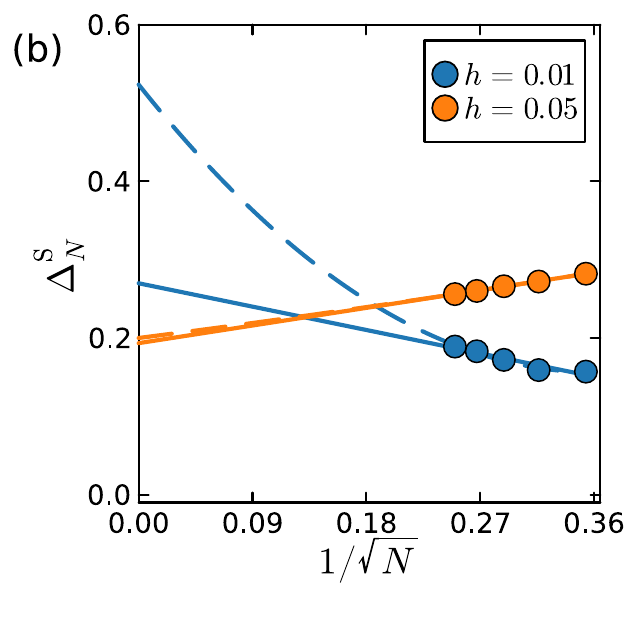}
\caption{The charge gap (a) and spin gap (b) extrapolated to linear (solid) and quadratic (dashed) order in $1/\sqrt{N}$ for a fixed $\lambda=1.1$. Similar analysis shows the charge gap persists at these values of $h$ across a broader range $0.7\leq\lambda\leq1.6$ (data not shown).}
\label{fig:charge gap}
\end{figure}

Before discussing the possibility of a first-order transition at small but finite $h$, we must first establish that the system on the paramagnetic side is indeed gapped. To this end, we examine the charge and spin gaps at $\lambda = 1.1$, a point within the paramagnetic phase but still close to the ferromagnetic boundary. As shown in \figref{fig:charge gap}, the system exhibits clear gaps in both sectors, confirming that the paramagnetic phase is gapped for $h \gtrsim 0.01$. Determining the precise value of $h$ at which the system transitions from the gapless CFL to a gapped phase is difficult and requires access to much larger system sizes, which are beyond the scope of this work. We have attempted to access larger system sizes using DMRG but found that it requires very large bond dimensions and it converges slowly in this regime, consistent with a large amount of entanglement in the CFL state. 

\begin{figure}[H]
\centering
\includegraphics[width = 0.95\textwidth]{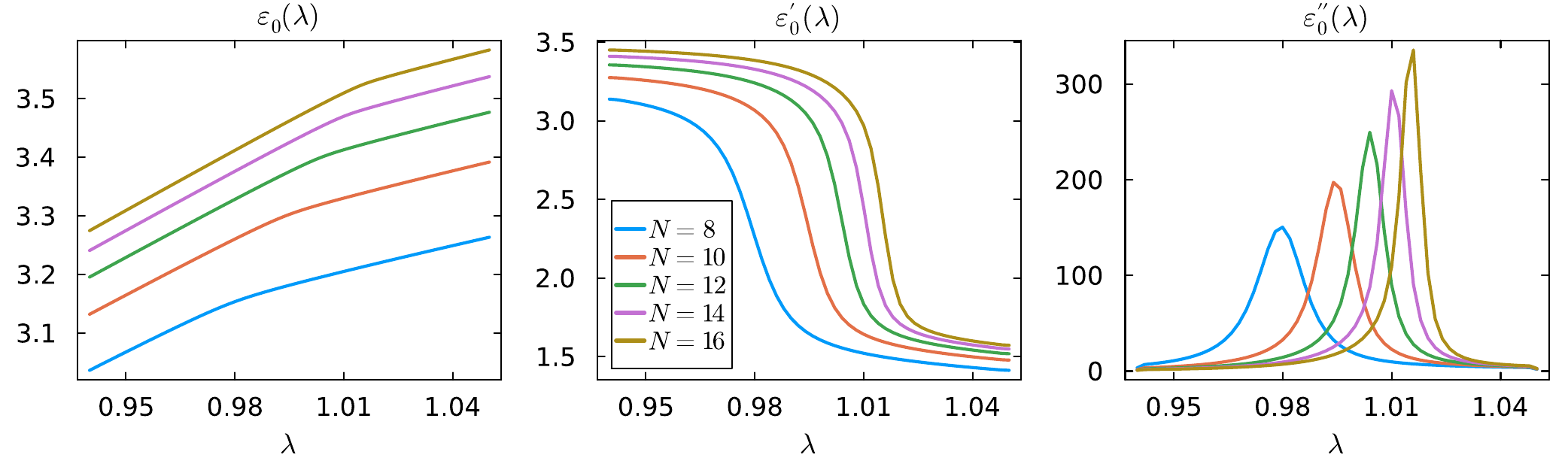}
\includegraphics[width = 0.95\textwidth]{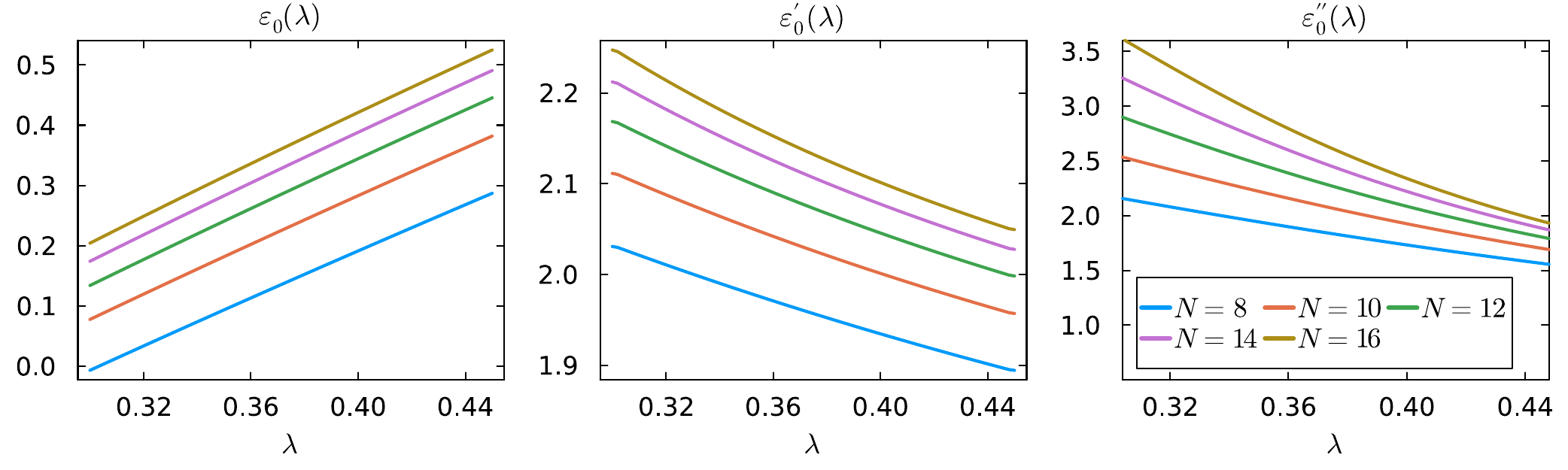}
\caption{The ground state energy density with its first and second derivatives with respect to $\lambda$. Top: Crossing the putative first-order transition at $h=0.05$. 
Bottom: Same as the top panel, but crossing the second-order Ising line at $h=2.0$. Note that $\lambda=0.4$ is where the finite size scaling curves cross in \figref{fig:TCI FSS}(c) of the main text and it is also where the spectrum in \figref{fig:phase diagram}(b) agrees with that of 3D Ising CFT. It is clear, both at the optimized conformal point $A$, $\lambda=0.35$ in \figref{fig:Ising Line}, and the finite-size crossing point, $\lambda\approx0.4$, the behavior is very smooth without any indication of a discontinuity, unlike in the first-order case shown in the top panel.
}
\label{fig:h_0.05_h_2.0_raw}
\end{figure}

Having established the presence of a gapped paramagnetic phase, we can now examine the putative first-order transition at small $h$. As discussed in the main text, we diagnose the transition by analyzing the ground state energy density, $\varepsilon_0 = E_0/N$, and its derivatives as a function of intralayer interaction $\lambda$. The results at $h = 0.05$ are shown in the top panel of \figref{fig:h_0.05_h_2.0_raw}. For comparison, we also compute the same quantities across the 3D Ising transition line, shown in the bottom panel of the same figure. While the Ising case exhibits smooth behavior in both the energy density and its derivatives, the small-$h$ case reveals a clear kink in the first derivative and a pronounced peak in the second derivative, both of which sharpen with increasing system size -- the hallmarks of a first-order transition.

\begin{figure}[t]
\centering
\includegraphics[width = 0.75\textwidth]{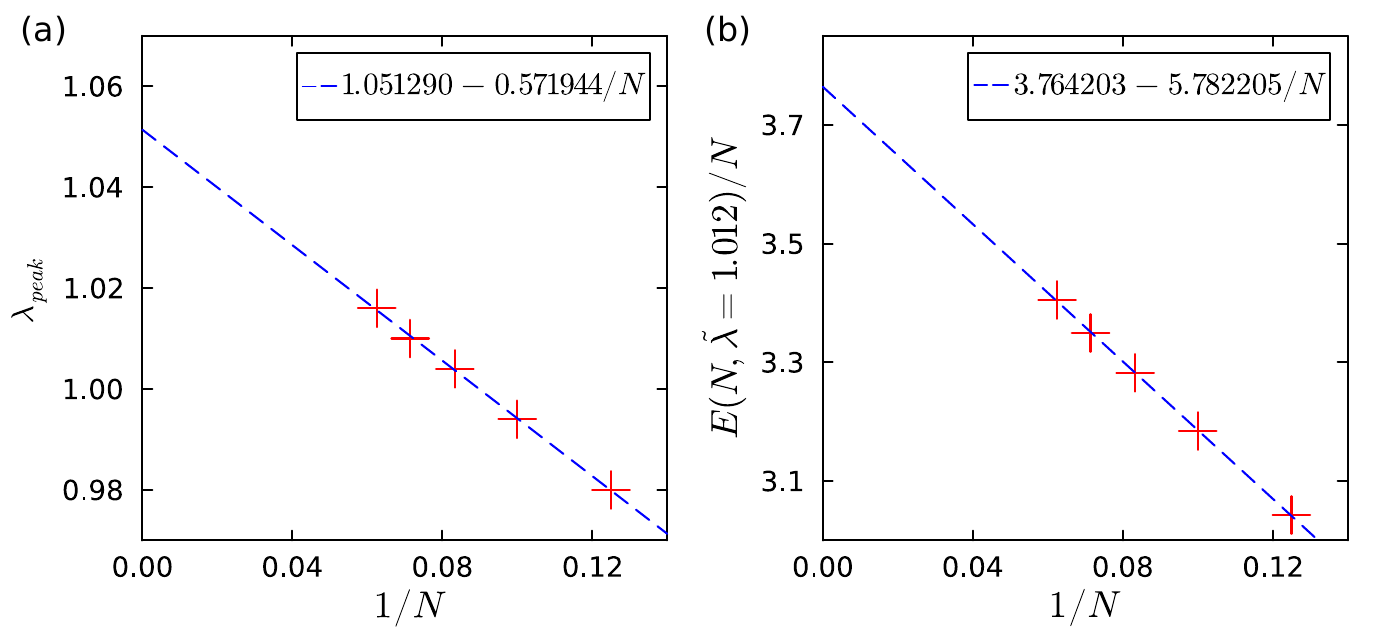}
\caption{Extrapolations on the energy density to verify  \eqnref{eqn:first_order}. (a) The location of $\lambda$ such that $\varepsilon_0^{\prime\prime}(\lambda)$ is maximized in \figref{fig:h_0.05_h_2.0_raw}. The linear coefficient in our extrapolation gives $b$ in \eqnref{eqn:first_order}. (b) Performing the appropriate rescaling of $\lambda$ using $b$, we fixed $\tilde{\lambda}=1.012$ and extrapolate the ground state energy density. The linear coefficient of this extrapolation represents $a_{-1}$.}
\label{fig:first_order_extraps}
\end{figure}

To better analyze the result across the first-order transition, we encounter one complication: the location of the peak keeps drifting with the system size, a feature that has been observed in various fuzzy sphere models~\cite{voinea2024regularizing,fan2025simulatingnonunitaryyangleeconformal}.
In order to better compare results across different system sizes, we adopt a particular ansatz for the scaling of the energy density with system size $N$, including the drift of the first order transition point. 
We separate the energy density into a singular and a smooth part:
\begin{equation}
\label{eqn:first_order}
    \varepsilon_0(N, \lambda) \equiv \frac{E(N,\lambda)}{N} = f_N(\lambda-\lambda_c - b/N) + a_0 + \frac{a_{-1}}{N} + a(\lambda - \lambda_c - b/N) + \ldots\,,
\end{equation}
where $f_N$ captures the singular behavior in the thermodynamic limit, $a_{-1},a_0,a$ and $b$ are constants. 
Namely, $b$ accounts for the drift of the transition point, and $a_{-1}$ captures the system size dependence of the energy density itself.
It is reasonable to assume that only $f_N(\lambda-\lambda_c - b/N)$ contributes the observed peak in the second order derivative of $\varepsilon_0$.
We can then estimate $b$ by fitting the $\lambda$ values at which the peak of $\varepsilon_0''$ occurs for each system size. Subsequently, recalculating the energy density by shifting $\lambda\rightarrow\lambda+b/N \equiv \tilde{\lambda}$, we can extrapolate the energy density in the ferromagnetic phase to extract an estimate for $a_{-1}$. These extrapolations are shown in \figref{fig:first_order_extraps}. 

\begin{figure}[H]
\centering
\includegraphics[width = 0.95\textwidth]{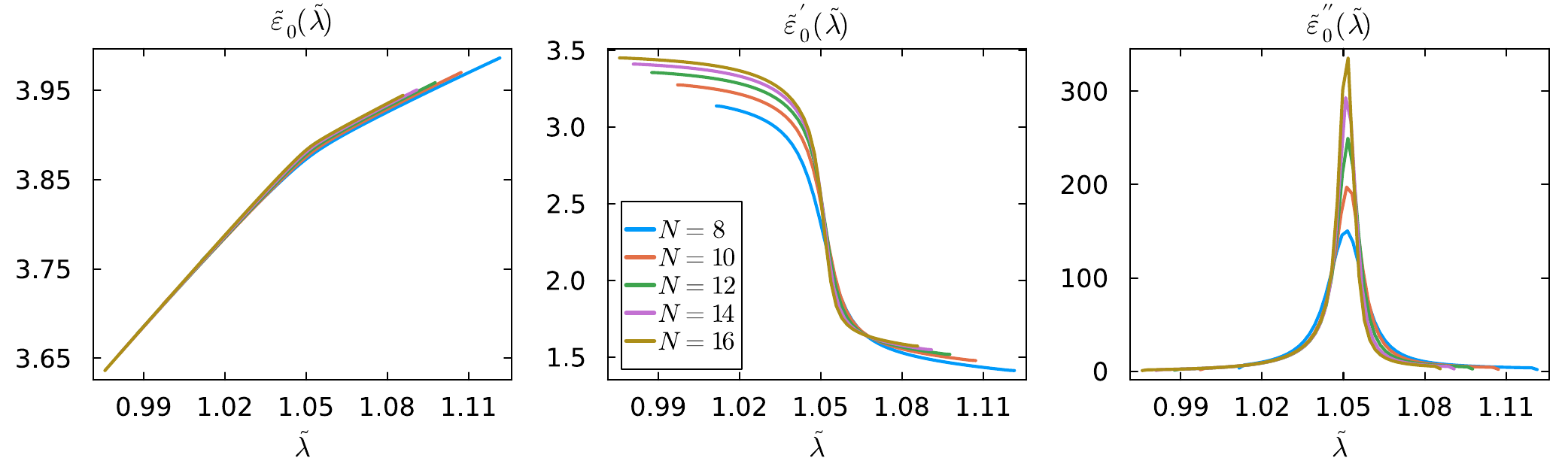}
\caption{The rescaled ground state energy density across a first order transition with $h=0.05$, along with its first and second derivatives with respect to $\tilde{\lambda} = \lambda+0.571944/N$.}
\label{fig:h_0.05_rescaled}
\end{figure}

Finally, using the acquired values of $b$ and $a_{-1}$, we can now plot the corrected ground state energy density with respect to these system size scaling coefficients. We define
\begin{equation}
    \tilde{\varepsilon}_0(\tilde{\lambda}) = \frac{E(N,\tilde{\lambda})}{N} - a_{-1}/N
\end{equation}
as our rescaled ground state energy density. The rescaled ground state energy density along with its first and second derivatives are shown in \figref{fig:h_0.05_rescaled}, which is the same as  \figref{fig:TCI FSS}(d) in the main text. 

\end{document}